\documentclass[10pt, aps,prc,twocolumn,superscriptaddress,preprintnumbers,
amsmath, 
floatfix,
longbibliography,
nofootinbib
]{revtex4-1}
\usepackage[T1]{fontenc}
\usepackage[utf8x]{inputenc} 
\usepackage{adjustbox}          % Used to adjust figure frames
\usepackage[caption=false]{subfig}
\usepackage{url}
\usepackage{color}
\usepackage{float}
\usepackage[pdftex,colorlinks=true, linkcolor = blue, citecolor=blue,urlcolor=blue, bookmarksnumbered=true, bookmarksopen=true]{hyperref}
\usepackage{longtable}
\usepackage{amsfonts}
\usepackage{dsfont}
\usepackage{wrapfig,bm} 
\usepackage[normalem]{ulem}
\usepackage{MnSymbol}

\newcommand{\beq}{\begin{equation}}
\newcommand{\eeq}{\end{equation}}
\newcommand{\bea}{\begin{eqnarray}}
\newcommand{\eea}{\end{eqnarray}}

\begin{document}

\title{Impact of the Center of Mass Fluctuations on the Ground State Properties of Nuclei}
  
\author{Matthew Kafker}
\affiliation{Department of Physics,  University of Washington, Seattle, Washington 98195--1560, USA}
\affiliation{Present affiliation: Cyclotron Institute , Texas A\&M University, College Station, TX 77843 USA}
\email{mmkafker.physics@gmail.com}
\author{Aurel Bulgac}% 
\email{bulgac@uw.edu}%
\affiliation{Department of Physics,  University of Washington, Seattle, Washington 98195--1560, USA}

\date{\today}

\begin{abstract}

Ground state properties across the entire nuclear chart are described predominantly and rather accurately 
within the  density functional theory (DFT).  DFT however  breaks  many symmetries, among them the 
most important being the translational, rotational, and gauge symmetries. 
The translational symmetry breaking is special, since it is broken for all nuclei, unlike the rotational
and  gauge symmetries. The center-of-mass (CoM) correction most commonly used in the literature 
[Vautherin and Brink, Phys. Rev. C {\bf 5}, 626 (1972) and Bender {\it et al.}, Rev. Mod. Phys. {\bf 75}, 121 (2003)]  
leads to a gain of 15,...,19 MeV, which varies
rather weakly for medium and heavy mass nuclei. 
A better approximation to the  CoM correction was suggested by Butler {\it et al.}, Nu cl. Phys. A {\bf 422}, 157 (1984)
and its magnitude varies between 10 and 5 MeV from light to heavy nuclei, a correction which is also significantly larger 
than the RMS energy error in the Bethe-Weizs\"acker mass formula, initially proposed by Gamow, Proc. Phys. Soc. A {\bf 126}, 157 (1930), 
which is at most 3.5 MeV, and which for heavy nuclei corresponds to about 0.2\% of their mass. 
The CoM energy correction due Butler {\it et al.}  is also significantly larger than the RMS energy deviation
achieved in any DFT evaluations of the nuclear masses performed without any symmetry restoration or 
zero-point energy fluctuations, with an energy RMS typically between 2 and 3 MeV.  
Here we analyze the CoM projection method suggested by 
Peierlsand Yoccoz, Proc. Phys. Soc. A {\bf 70}, 381 (1957), which leads to a translationally invariant many-body wave function, in 
a procedure fully equivalent to those suggested for restoring rotational and gauge symmetries. 
This is the only approach for the evaluation of the CoM energy correction to the mean field binding energies, 
which is not contaminated by contributions from excited states, which  leads to estimates of the CoM 
energy corrections consistently superior in quality to those used routinely in literature. We have also outlined 
how the nuclear EDF needs to be upgraded to include relativist effects, in order to correctly reproduce the 
observed nuclear masses. 

\end{abstract} 

\preprint{NT@UW-25-5}

\maketitle  

\section{Introduction} \label{sec:I}

Within the mean field approximation it is desirable to evaluate the correction to the total energy of a nucleus
due to the CoM fluctuations. In any mean field treatment of many-fermion systems the CoM coordinate and momentum 
have well defined values, even though the corresponding operators for the CoM coordinate and momentum do not commute. 
By definition the mean field, defined as an expectation value over a mean field wave functions of a many-body operator 
in general, has also a well defined value with absent by definition quantum fluctuations.
In nuclear physics in particular the Energy Density Functional (EDF) is constructed to satisfy a generalization of the 
Galilean invariance under local gauge transformations~\cite{Engel:1975,Bender:2003,Bulgac:2018,Bulgac:2007}. 
This implies that for any isolated small volume of a system one can uniquely separate the CoM and the 
intrinsic energies of that volume. In time-dependent treatment this is a crucial aspect, which for example 
in the case of a reaction can identify exactly the intrinsic energies of any reaction fragments, irrespective of the 
velocity any of those fragments might move at any given moment.  Long-range forces, in particular the 
Coulomb interaction between reaction fragments even at large separations, can lead to the time-dependent 
polarization of reaction fragments and thus can induce relatively small amplitude currents however. 
In static mean field treatments is  often of interest to correctly evaluate the role of CoM fluctuations 
on the total energy of the system, as these fluctuations can and do affect the expectation 
values of many observables of interest.
 
The following CoM energy correction to the ground state mean field energy of a 
nucleus of mass $A$ is considered by most authors to be themes reliable 
\begin{align} 
&{E}_{\rm CoM} =- \left \langle \psi |   \hat{\rm K}_{\rm CoM}| \psi \right  \rangle  =-\left \langle \psi \left | \frac{ \hat{\bf P}^2_{\rm CoM} }{2Am} \right |  \psi \right  \rangle \nonumber \\
&=-\left \langle \psi \left |  \sum_i\frac{\hat {\bf p}^2_i}{2Am}+\sum_{i<j} \frac{\hat{\bf p}_i\cdot\hat{\bf{p}}_j}{Am} \right |  \psi \right  \rangle <0 ,
\label{eq:KE0}
\end{align}
where $|\psi \rangle$ is the mean field many-body wave functions, 
$\hat{\bf P}_{\rm CoM}=\sum_i\hat{\bf p}_i$ and $m\approx (m_P+m_N)/2$ is the average nucleon mass.  \footnote{Some  EDFs  in literature use an
\emph{ad hoc} prescription, and 
include implicitly CoM energy correction to the ground state energy in their parameterization, while other authors  do not. 
The CoM energy correction decays quite fast with mass number A, and vanishes for infinite systems, 
and  does not have a mass dependence similar to the Bethe-Weizs\"{a}cker 
mass formula or to the various contribution routinely included in EDFs, 
such as volume ($\propto A$), surface ($\propto A^{2/3}$), and Coulomb ($\propto Z^2/A^{1/3}$).
Gogny type of EDF often include the contribution CoM kinetic energy correction exactly.}
A simpler approximation is to renormalize 
the nucleon mass according to the widely used prescription~\cite{Vautherin:1972,Bender:2003,Ring:2004}
\begin{align}
\sum_i\frac {\hat{\bf p}_i^2 }{2m} \rightarrow  \sum_i \frac {\hat{\bf p}_i^2 }{2m} \left (1 -\frac{1}{A}\right ), \label{eq:HF}
\end{align}
which, neglecting surface effects, leads to  
\begin{align}
E_{\rm CoM} \approx  -\frac{1}{A} \left \langle \psi \left |\sum_i \frac{ \hat{\bf p}_i^2 }{2m} \right |  \psi \right  \rangle
\approx -\frac{3}{5} \varepsilon_F \approx -22 \,{\rm MeV}, \label{eq:KE1}
\end{align} 
using $\varepsilon_F = \hbar^2k_F^2/2m$ and nuclear saturation density $n_0=2k_F^3/3\pi^2 = 0.16$ fm$^{-3}$.
Thus $E_{\rm CoM}$  is largely independent of the 
nuclear mass and arguably larger than any other corrections arising 
from the restoration of either rotational, parity, isospin, or gauge symmetries.
Here $\varepsilon_F$ is the Fermi energy of symmetric nuclear matter in its ground state.  
The actual numbers for $E_{\rm CoM}$  evaluated with Eq.~\eqref{eq:HF} and the SeaLL1 
EDF~\cite{Bulgac:2018} (overlooking isospin and Coulomb corrections)
are approximately reproduced with 
\begin{align}
E_{\rm CoM}^{\rm HF}= -17.46 - \frac{15.17}{ A^{1/6}} + \frac{28.01}{ A^{2/6} }\,\pm 0.42 \,[{\rm MeV}], \label{eq:ehf}
\end{align} 
with some noticeable discrepancies only for Ca isotopes. 

The rotational energy correction was  evaluated using different approximations, see Ref.~\cite{Bender:2003}.  For example,  \textcite{Scamps:2021} 
obtain a rotational energy correction with a maximum value of 4.07 MeV and a mean value of 1.69 MeV, and with a RMS of 0.76 MeV, using a formula very 
similar to Eq.~\eqref{eq:KE0}, but for rotations around principal axes and using Belyaev moments of inertia. Experimental data  that the use of such 
phenomenological formula is appropriate for the ground state band up to not very high angular momenta, though to what extent the rotation 
is uniform, with constant angular velocity, was never shown in a fully microscopic derivation. In case of quantized vortices in neutron matter
found in the crust of neutron stars it was clearly shown that the rotation of the neutron superfluid is not uniform\cite{Yu:2003a}.  \textcite{Bender:2006}  
obtain similar values for the rotational energy correction, a maximum value of 4.2 MeV and a mean value of 2.66 MeV, and  with an RMS of 0.51 MeV, using a 
combination of the generator coordinate method (GCM) for quadrupole correlations and the angular momentum projection.  
Similar results were obtained for correlation energy correction using the Gogny effective forces (implicitly including the CoM energy correction as well) 
by ~\textcite{Delaroche:2010}, with a minimum value -7.71 MeV, maximum value +4.44 MeV, mean value  -2.84 MeV and an RMS of 1.39 MeV.

\begin{figure}
\includegraphics[width=0.9\columnwidth]{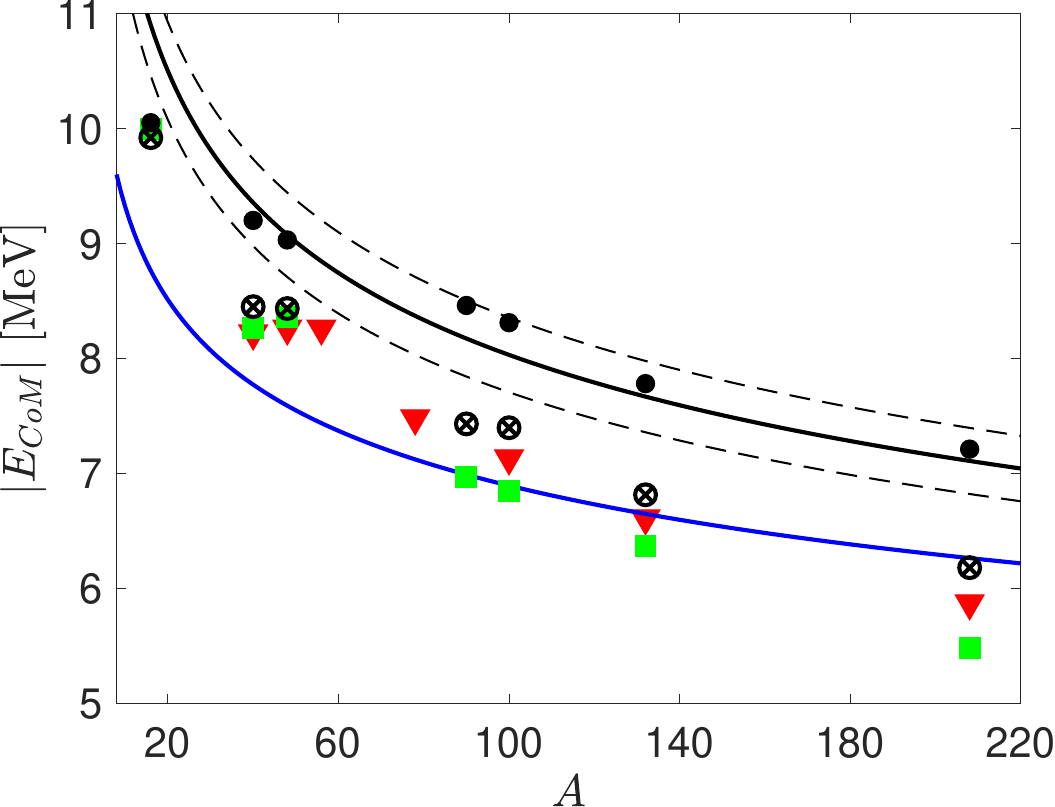}
\caption { \label{fig:ecom}   The most accurate estimates of  $|E_{\rm CoM}|$ reported so far in literature, along with the results of this work. 
The black lines and dots (solid line for average and dashed lines for upper and lower for error bars) are the results of the present work, 
where we evaluated the PY CoM energy correction
with Eq.~\eqref{eq:com}, $E_{\rm CoM}  =  \langle \Psi_{\bm 0}| \hat{\rm H} | \Psi_{\bm 0} \rangle -
\langle\psi |\hat{\rm H}|\psi \rangle= E_{gs}-E_{MF}\leq 0$, where $|\psi\rangle$ and $|\Psi_{\bf 0}\rangle$ are the mean field and the PY projected
many-body wave functions respectively.  The black circles with crosses are 
our evaluation of $E_{\rm CoM}=- \left \langle \psi |   \hat{\rm K}_{\rm CoM}| \psi \right  \rangle $ using the alternative expression Eq.~\eqref{eq:MF11} 
and identical to Eq.~\eqref{eq:KE0},  in Section~\ref{sec:II}, but using the PY framework.
The red triangles represent the results of \textcite{Da-Costa:2024} for $E_{\rm CoM}$ using Eq.~\eqref{eq:KE0}.  
The black circles with crosses and the red triangles are surprisingly similar in values, even though they were obtained with different EDFs.
The blue line represents  \textcite{Dobaczewski:2009}'s results using a combination of the PY framework and the Lipkin translational symmetry restoration prescription.  
We used  SeaLL1 EDF~\cite{Bulgac:2018} for the estimates for $E_{\rm CoM}^{\rm HF}$, see Eq.~\eqref{eq:ehf}, 
to generate \textcite{Butler:1984} estimates with green squares, see also Eq~\eqref{eq:butler}. This plot clearly demonstrates 
that the PY framework leads to larger CoM energy correction that the usual formula 
${E}_{\rm CoM} =- \left \langle \psi |   \hat{\rm K}_{\rm CoM}| \psi \right  \rangle $, see Eq.~\eqref{eq:KE0}.}
\end{figure}

Since the many-body wave functions used in Eqs.~(\ref{eq:KE0}, \ref{eq:KE1}) are not translational invariant,
these particular CoM energy corrections are contaminated by contributions from nucleus excited states, 
as we show in this work, see Section~\ref{sec:II}. This is particularly important, since the 
restoration of the translational invariance is relevant for all nuclei, while the broken symmetries are restored exactly
for a fraction of nuclei only~\cite{Bender:2003,Sheikh:2021}. Berger {\it et al.}~\cite{Berger:1984} 
have included the $E_{\rm CoM}$ correction Eq.~\eqref{eq:KE0}, which affects the  fission barriers 
with Gogny interaction DS1, as also the results of Ref.~\cite{Da-Costa:2024} suggest, 
where the importance of this correction and in particular its impact on nuclear surface energy is discussed.
The results for $E_{\rm CoM}$ from  Refs.~\cite{Da-Costa:2024,Dobaczewski:2009} reproduced in Fig.~\ref{fig:ecom} were obtained 
using the exact expression for $\hat{\bf P}_{\rm CoM}$. 

It is important to distinguish between binding energies 
evaluated in a pure EDF framework~\cite {Vautherin:1972,Bender:2003,Bulgac:2018}, where there are no corrections 
due to symmetry restoration (except maybe 
Eq.~\eqref{eq:HF} with RMS error ${\cal O}(3)$ MeV and those evaluated by the group of 
Goriely~\cite{Scamps:2021,Grams:2023,Da-Costa:2024} (see also earlier papers), with RMS errors $\approx 0.5\ldots 0.7$ MeV,
where the corrections due to symmetry restoration and other effects 
are included and used to correct the EDF parameters. 
As typical in evaluating the nuclear binding energies  in several dozens of studies performed 
by Goriely's group over the last few decades,  these authors do not routinely
publish the magnitudes of various corrections arising from 
restoring various symmetries, but only the final cumulative results obtained after the contributions from the 
restoration of all symmetries chosen by these authors have been accounted for and the 
EDF parameters have been refitted. 
A related aspect is the evaluation of the CoM energy correction 
in the studies, which are performed using an expression derived from Skyrme-like EDF. It was shown by \textcite{Negele:1972} 
that an effective coordinate dependent mass appears when non-local mean fields are expressed in an Density Matrix Expansion (DME). Thus, the difference 
between CoM energy corrections using an inverse nucleon mass and an inverse effective mass is due to nucleon interactions, while $E_{\rm CoM}$, 
see Eq.~\eqref{eq:KE0} is strictly the expectation of the CoM kinetic energy alone. Since inside a nucleus the values used 
are  $m_{\mathit{eff}}\approx 0.80-0.85 m$, the reported values for the CoM energy corrections evaluated in these works are enhanced 
by a factor of $\approx 1.2$, due to the role of the nucleon-nucleon interactions,  with respect to the pure 
CoM kinetic energy expectation value obtained with bare nucleon masses in  a pure mean field. The role of internal motion and interactions 
on the mass defect is a relativistic effect in determining the rest mass of a bound system of particles, see  discussion of the end of this Section. \textcite{Engel:1975} have 
shown that one should consider the combination $\tau({\bf r})n({\bf r})- {\bf j}^2({\bf r})$ in the interaction part of the EDF, 
where $\hbar^2\tau({\bf r})/2m$ is the kinetic energy density,  ${\bf j}({\bf r})$ is the local current density and 
$n({\bf r})$ is the nucleon number density. Only then the total energy of a system 
are invariant under a Galilean boost and the expected nonrelativistic mass of the system is indeed $Am$.

It was argued by \textcite{Butler:1984}, see also \cite{Bender:2003},
that a better estimate of the CoM energy correction than Eq.~\eqref{eq:HF} is given by a different prescription
\begin{align}
&\sum_i \frac {\hat{\bf p}_i^2 }{2m} \rightarrow \sum_i \frac {\hat{\bf p}_i^2 }{2m} \left (1 -\frac{2}{(N+2)A}\right ),  \label{eq:But}\\
&A = \frac{2}{3} [ (N+2)^3-(N+2)].
\end{align} 
$N$ is typically the number of fully occupied harmonic oscillator shells  in a closed-shell nucleus,
with $N=Z$. \textcite{Butler:1984}'s prescription leads $E_{\rm CoM}$ which varies from 12.20 MeV for $^{16}$O to 5.19 MeV for $^{208}$Pb, 
with a mass dependence 
\begin{align}
&E_{\rm CoM}\approx -\frac{2}{A(N+2)} \left \langle \psi \left |\sum_i \frac{ \hat{\bf p}_i^2 }{2m} \right |  \psi \right  \rangle 
\approx - \frac{6}{5(N+2)}\varepsilon_F, \label{eq:butler}
\end{align}
The difference between the Hartree-Fock (HF) $E_{\rm CoM}$  using Eq.~\eqref{eq:HF} 
$E_{\rm CoM}\approx 19$ MeV for $^{208}$Pb and   $E_{\rm CoM}\approx 5.2 $ MeV using Eq.~\eqref{eq:But},
is unexpectedly large,  comparable to the difference between the CoM estimates obtained with Eq.~\eqref{eq:KE0} and Eq.~\eqref{eq:HF}, which is 
significantly larger than the RMS error in the binding energy in the Bethe-Weizs\"{a}cker mass formula. With
the \textcite{Butler:1984} correction the RMS for binding of spherical nuclei alone dropped significantly from 1.54 to 0.97 MeV in case of SeaLL1 EDF~\cite{Bulgac:2018}. 
 
In Fig.~\ref{fig:ecom} we compare the most current and best
estimates of $|E_{\rm CoM}|$ for the magnitude and its $A$-dependence, obtained by various authors, including present results.
Considering all the studies performed in the past, one can state that up to this day there is no convincing study 
or even agreement between different claims. 
\begin{itemize}
\item  With red triangles in Fig.~\ref{fig:ecom} are the results reported by \textcite{Da-Costa:2024}, Fig. 9 obtained with 
${E}_{\rm CoM} =- \left \langle \psi |   \hat{\rm K}_{\rm CoM}| \psi \right  \rangle$, see Eq.~\eqref{eq:KE0}.
\item  The blue line in Fig.~\ref{fig:ecom}  are the results $E_{\rm CoM}\approx 12.6(8) A^{-0.131(11)}$, reported  by \textcite{Dobaczewski:2009} 
in Fig. (4), due to zero-point CoM fluctuations.
\item   The green squares in Fig.~\ref{fig:ecom} are \textcite{Butler:1984} results, see Eqs.~(\ref{eq:butler}) evaluated here with SeaLL1 EDF. 
In Ref.~\cite{Bulgac:2018}, where the non-optimized SeaLL1 EDF was introduced, after adding the \textcite{Butler:1984} CoM energy corrections  to
the binding energy RMS of even-even spherical nuclei decreased from 1.54 MeV to 0.97 MeV, by correcting the ground state 
energies of light and medium mass nuclei predominantly.
\item  $E_{\rm CoM}\propto A^{-1}$ (not shown) according to the arguments presented 
by \textcite{Sheikh:2021} in Section 3.2.2, where the authors discuss the restoration of ``Linear momentum.''  
The $E_{\rm CoM}\propto A^{-1}$ is clearly never reproduced by other estimates. If this prediction from Ref.~\cite{Sheikh:2021} 
would be correct the CoM energy correction in $^{208}$Pb would be more than 5 times smaller than in $^{40}$Ca.   
\item With black circles with crosses are our results for ${E}_{\rm CoM} = \left \langle \psi |   \hat{\rm K}_{\rm CoM}| \psi \right  \rangle$, see Eq.~\eqref{eq:MF11}.
\item The black lines and black dots are our CoM energy corrections obtained with translational invariant 
many-body wave function $\Psi_{\bm 0}$, see Eq.~\eqref{eq:wf}~\cite{Peierls:1957}, with 
$E_{\rm CoM}  =  \langle \Psi_{\bm 0}| \hat{\rm H} | \Psi_{\bm 0} \rangle- 
\langle\psi |\hat{\rm H}|\psi \rangle$,  Eq.~\eqref{eq:com} reported in this work.
\end{itemize}
Equally, it is not clear how the magnitude of $E_{\rm CoM}$ compares in importance 
with any other corrections from other required symmetry restorations.  The simple recipe used by \textcite{Vautherin:1972}, see Eqs.~(\ref{eq:KE1}, \ref{eq:ehf}) 
point to a finite (thus incorrect) asymptotic value for $E_{\rm CoM}\approx 18$ MeV,  \textcite{Da-Costa:2024} using 
Eq.~\eqref{eq:KE0} report a different behavior, $E_{\rm CoM}\approx 6\ldots 8$ MeV for all magic nuclei, unlike \textcite{Butler:1984}, see Eq.~\eqref{eq:butler}, 
and  in clear disagreement with  $E_{\rm CoM}$ reported in Ref.~\cite{Dobaczewski:2009}, 
and in a stark disagreement with the claims made in Ref.~\cite{Sheikh:2021}.  

Current mean field frameworks do not provide accurate enough estimates for the nuclear binding energy (BE) 
needed to predict nuclear abundances and reactions in stars according to a number of authors. 
Even after including a plethora of beyond mean field (BMF) corrections 
the needed accuracy in not sufficient, but that conclusion depends on the study, see 
Refs.~\cite{Schatz:1998,Schatz:2001,Mumpower:2016,Cowan:2021,Da-Costa:2024} and references therein, where it is claimed that
an RMS error not greater than ${\cal O}(100)$ keV in BE is needed when performing the simulation of the $rp$-process nuclear reaction network. 
In other studies however, see Ref.~\cite{Grams:2021}, agreements with observed abundance in the solar system are obtained 
with current EDFs when evaluating the $r$-process~\cite{Lemaitre:2021}.
CoM energy corrections also appear to play a particularly  important role in determining the nuclear surface tension 
and correspondingly the heights of fission barriers~\cite{Berger:1984,Da-Costa:2024}. 

\textcite{Peierls:1957,Peierls:1962} made the claim that in order to restore Galilean invariance the nucleus mass 
should be  $M_A= Am$, a claim which can be found in literature until recent times~\cite{Ring:2004,Dobaczewski:2009,Sheikh:2021},
a statement formally correct only in the limit $c\rightarrow \infty$.  
It is known since the beginning of the 20th century that only the experimentally measured (physical) mass of a nucleus 
$M_A \neq Am$  determines its correct CoM kinetic energy
\begin{align} 
&\hat{\rm K}_{\rm CoM}=\frac{\hat{\bf P}^2_{\rm CoM}}{2M_A},\quad {\rm where} \quad \underline{M_A c^2= Amc^2-{\rm BE}}, \label{eq:PY_mass}\\
& \hat{\bf P}_{\rm CoM}=\frac{1}{A}\sum_i \hat{\bf p}_i, \quad {\bf R}_{\rm CoM}= \frac{1}{A}\sum_i{\bf r}_i, \label{eq:p_com}
\end{align}  
and where the binding energy ${\rm BE}$ is a relativistic correction of ${\cal O}(0.01\,Amc^2)\approx 7\ldots 8 \times A$ MeV.
Only this kinetic energy can be measured and confronted with theory in the non-relativistic limit, when ${\rm P}_{\rm CoM} \ll M_A c$.
The PY nuclear masses reported  Ref.~\cite{Dobaczewski:2009,Sheikh:2021}, are 
in clear disagreement with $M_Ac^2 =Amc^2 - BE$~\eqref{eq:PY_mass} with BE evaluated in either mean field, mean 
field with corrections, or with BE extracted from measured nuclear masses. In the case of $^{208}$Pb the correction to 
the evaluated mass reported in Refs.~\cite{Sheikh:2021,Dobaczewski:2009} is $\approx 10\ldots 15 $ GeV, 
almost an order of magnitude larger than the experimental measured mass BE$(^{208}$Pb) = 1.635 GeV. 

In all non-relativistic DFT implementations  it is assumed that nuclear masses satisfy
\begin{align} 
&\frac{M_A({\rm exp}) }{N \,m_n+Z\, m_p^2 }\equiv 1, \label{eq:inertia0}\\
\end{align}
where $m_{n,p}, \, M_A({\rm exp})$ are the measured neutron, proton and nuclear masses respectively,
a statement which goes all the way back to \textcite{Peierls:1957}  and it is often related to 
Galilean invariance~\cite{Schmid:2004,Bender:2003,Dobaczewski:2009,Sheikh:2021}, and not 
merely to Galilean translational invariant wave functions. 
While the Galilean invariance is built in the Schr\"odinger equation and in non-relativistic DFT,
 any predicted nuclear inertia are in strong violation with Nature, since 
\begin{align} 
&\frac{M_A ({\rm DFT})\,c^2}{N \,m_n\,c^2+Z\, m_p^2\,c^2 }< 1, \label{eq:inertia1}\\
&\frac{ M_{^{208}Pb }( {\rm SeaLL1})}{N \,m_n+Z\, m_p^2 } =0.991624,
\end{align} 
thus a violation of Galilean invariance of $\approx 1$\%. \textcite{Dobaczewski:2009} in 
his Peierls  \& Yoccoz's implementation of translational-symmetry restoration obtains
\begin{align} 
&\frac{M_A ({\rm DFT})\,c^2}{N \,m_n\,c^2+Z\, m_p^2\,c^2 } = 1.311(14)\,A^{-0.060(3)}, \label{eq:inertiaD}
\end{align}
with values $\approx 1.10$ for $A=16$  and $\approx 0.95$ for $A=208$, thus 
with significant consequences on many observables.

\section{Peierls and Yoccoz's translational symmetry restoration method} \label{sec:II}

In this work we will use the method suggested a long time ago by 
\textcite{Peierls:1957} (PY), and which has been  used rather scarcely over the years 
and only for nuclei with $A\leq 40$ in 
Refs.~\cite{Marcos:1984,Schmid:1990,Schmid:1991,Schmid:2002a,Rodriguez-Guzman:2004a,Schmid:2004}, 
with the exception of Ref.~\cite{Dobaczewski:2009}. 
Ref.~\cite{Rodriguez-Guzman:2004a} is the latest and most complete study of CoM projection using the PY 
procedure. However, these authors only evaluate the differences between the mean field evaluation of the total 
energy of nuclei with $\hat{\rm T}_{\rm CoM}$ subtracted according to Eq.~\eqref{eq:KE0} 
and the results of the implementation of the PY procedure are a variation-after-projection (VAP), 
so we cannot compare the total magnitude of the CoM correction obtained by these authors with our results. 
It is however interesting to learn from their results that  the difference between the projection-after-variation 
(the method used in this work) and VAP method is  about 1.5 MeV or less, 
noticeably smaller than the entire $E_{\rm CoM}$ as evaluated by us, but still important to account for.

PY  suggested a straightforward center-of-mass projection for an arbitrary 
many-body wave function   $\psi({\bf r}_1,\ldots,{\bf r}_A)$ for $A$ nucleons
\begin{align}
& C_{\bm n} \Psi_{\bm n}({\bf r}_1,\ldots,{\bf r}_{A})=  \nonumber \\
& \int \!\!\!\frac{ d^3{\bf a}}{V} \psi({\bf r}_1-{\bf a},\ldots,{\bf r}_A-{\bf a}) e^{ - i  {\bf P}_{\bm n}\cdot ( {\bf R}-{\bf a} )/\hbar }, \label{eq:wf}\\
&\psi({\bf r}_1,\ldots,{\bf r}_{A})=
\sumint_{\bm n}C_{\bm n}\Psi_{\bm n}({\bf r}_1,\ldots,{\bf r}_{A}) e^{  i {\bf P}_{\bm n}\cdot {\bf R}/\hbar }. \label{eq:PY_wf}
\end{align}
$|C_{\bf n}|^2$ is the weight of 
finding the component with the CoM ${\bf P}_{\bm n}$ in  $\psi({\bf r}_1,\ldots,{\bf r}_A)$,
$V$ is the volume of the simulation box, and $\langle \psi|\psi\rangle =\langle \Psi_{\bm  n}|\Psi_{\bm n}\rangle =1$.
The factor $1/V$ is required to ensure the operation defined by Eq.~\eqref{eq:wf} is a projector.
We do not explicitly display the spin and isospin coordinates, whose presence does not affect the argument. We included the factor $1/(2\pi)^3$  
where needed in the definition of the coefficients of this type of expansion. 

The intrinsic dynamics is characterized by a well defined CoM momentum ${\bf P}_{\bm n} = \tfrac{2\pi {\bm n}\hbar}{L}$, 
where  ${\bm n}=(n_x,n_y,n_z)$ are 3D vectors with integer coordinates and $L$ is a cubic box of side length $L$.  
The location of the nucleus in the box of a finite but sufficiently large size is chosen so the unprojected wave 
function $\psi({\bf r}_1,\ldots,{\bf r}_{A})$ is negligible at the boundary, which can always be
achieved for  a bound many-body mean field state. $L$ and $\tfrac{L}{2n}$ define 
the infrared and ultraviolet wave lengths of the spectrum, which physically are determined by 
the size of the Universe and Planck's length scale ultimately, thus never infinitely large or small.  

The CoM projected many-body wave functions $\Psi_{\bm n}({\bf r}_1,\ldots,{\bf r}_{A})$ depend actually on only $A-1$ independent 3D coordinates, 
unlike the function $\psi({\bf r}_1,\ldots,{\bf r}_{A})$, as by construction 
\begin{align}
& \Psi_{\bm n}({\bf r}_1,\ldots,{\bf r}_{A})\equiv  \Psi_{\bm n}({\bf r}_1+{\bf a},\ldots,{\bf r}_{A}+{\bf a}), \quad \forall \,{\bf n},   \label{eq:Psi1}\\
& \underline{\hat{\bf P}_{\rm CoM} \Psi_{\bm n}({\bf r}_1,\ldots,{\bf r}_{A}) \equiv {\bf 0}}, \quad \forall \,{\bf n}, \label{eq:Psi2}
\end{align}
where the 3D vector ${\bf a}$ is arbitrary vector and  $\hat{\bf P}_{\rm CoM}$ is defined in Eq.~\eqref{eq:p_com}, while
\begin{align}
& \psi({\bf r}_1,\ldots,{\bf r}_{A})\neq  \psi({\bf r}_1+{\bf a},\ldots,{\bf r}_{A}+{\bf a}), \\
& \hat{\bf P}_{\rm CoM} \psi({\bf r}_1,\ldots,{\bf r}_{A}) \neq{\bf 0}.
 \end{align} 
 This property of the projected intrinsic many-body wave functions leads (somewhat unexpectedly) to the conclusion that 
 \begin{align}
 &\underline{\langle \Psi_{\bf n}| \,|{\bf P}_{\bf n}|^2|\Psi_{\bf n}\rangle { \equiv 0}}, \quad \forall \,\, {\bf n}. \label{eq:ec}
 \end{align}
 
 Since the single-particle basis set of wave functions in a box are plain waves, 
 there is a very simple way to illustrate Eqs.~(\ref{eq:Psi1}- \ref{eq:Psi2}). A many-body basis set 
wave function in that case  separates naturally into the CoM and intrinsic parts
\begin{align}
& \psi_{\bf P}( {\bf r}_1,\ldots,{\bf r}_A )= \exp\left ( i\sum_{ k=1}^A \frac{ {\bf p}\cdot {\bf r}_k }{ \hbar }  \right ) \nonumber\\
&= \exp\left ( i\frac{ {\bf P}\cdot{\bf R} }{\hbar}\right ) \exp \left ( i\sum_{k=1}^A \frac{ {\bf p}_k\cdot ({\bf r}_k-{\bf R}) }{\hbar}  \right),\\
&\Psi_{\bf P}({\bf r}_1,\ldots,{\bf r}_{A}) = \exp \left ( i \sum_{k=1}^A  \frac{ {\bf P}\cdot ({\bf r}_k-{\bf R} ) }{\hbar}  \right ),\\
&\sum_{k=1}^A ({\bf r}_k-{\bf R})\equiv {\bf 0}, \,\,\, \hat{\bf P}_{\rm CoM} \Psi_{\bf P}({\bf r}_1,\ldots,{\bf r}_{A}) \equiv {\bf 0}.
\end{align}
 
 In literature a number of authors have claimed that the true Galilean invariance is not restored by the 
 PY projection~\cite{Peierls:1957,Peierls:1962,Dobaczewski:2009,Sheikh:2021} and 
 that the kinetic energy of the CoM of a nucleus moving at constant speed is not given by 
 the expected non-relativistic expression 
 \begin{align}
{\rm K}_{\rm CoM}=\frac{{\bf P}^2_{\rm CoM}}{2Am},
\end{align} 
where ${\bf P}_{\rm CoM}$ is the expectation of the CoM momentum, see also Eq.~\eqref{eq:MF11} below, 
 and only the translational invariance is restored by the PY projection of the CoM.
 
It is obvious however that the total PY projected many-body wave function of a nucleus boosted with CoM momentum ${\bf P}_{\bf n}$ 
(in a simulation box with periodic boundary, where all momenta are quantized)
and it corresponding CoM kinetic energy~\cite{Shi:2020} are given by
\begin{align}
&\tilde{\Psi}_{\bf n}({\bf r}_1,\ldots,{\bf r}_{A})=\exp\left ( i\frac{ {\bf P}_{\bf n}\cdot {\bf R} }{\hbar} \right ) \Psi_{\bm n}({\bf r}_1,\ldots,{\bf r}_{A}),\\
&\langle \tilde{\Psi}_{\bf n}| \hat{\rm K}_{\rm CoM}| \tilde{\Psi}_{\bf n}\rangle =\frac{ {\bf P}^2_{\bf n} }{2Am},\,
E_{\bf n}=\langle \tilde{\Psi}_{\bf n}|\hat{\rm H}- \hat{\rm K}_{\rm CoM}| \tilde{\Psi}_{\bf n}\rangle ,
\end{align}
if the EDF satisfies a generalization of the Galilean invariance under local gauge 
transformations~\cite{Engel:1975,Bender:2003,Bulgac:2018,Bulgac:2007}. Our argument agrees with the argument made by \textcite{Ring:2004}.
We show in Section~\ref{sec:IV} how a relativistic DFT can be formulated and which will also lead as a result to a correct nucleus translational 
inertia is all agreement with experiment, see Eq. (8).

Even though the validity of the non-relativistic Galilean invariance in the PY framework is theoretically satisfied and 
the inertial mass of a nucleus emerges as expected in the CoM PY projection, this aspect has no relevance to reality. 
The measurable inertial mass of any reaction nuclear fragment differs from the naive Galilean expectation that  $M_A = Am$ by almost 1\%, 
due to relativistic effects, see Eq.~\eqref{eq:PY_mass}. The kinetic energy of nuclear fragments it typically measured with a
significantly higher accuracy than that and clearly, the presently used  in literature DFT approaches do not reproduce this aspect.

%%%%%%%%%%%%%%%%%%%%%%%%%%%%%%%%%%%%%%%%%%%%%%%%%%%%%%%%%%%%%%%%%%
There is a very simple example illustrating that all functions  $\Psi_{\bm n}({\bf r}_1,\ldots,{\bf r}_{A})$ 
do not depend on the CoM coordinate, but only of relative particle coordinates. 
Consider two particles in 1D with the corresponding wave
functions defined above $\psi(x,y)$ and $\Psi_P(x-y)$,   here for simplicity assuming that $\hbar =1$ and $R =\tfrac{x+y}{2}$,
\begin{align}
\psi(x,y) &= \sumint_{P,k} c_{P,k} \exp ( iPR + ik(x-y) ) \nonumber\\
 &=\sumint_P \exp( iPR )C_P\Psi_P(x-y),\label{eq:2}\\ 
C_P \Psi_P(x-y) &=\sumint_R \exp(-iPR) \psi(x,y)\nonumber \\
& = \sumint_k c_{P, k}\exp(ik(x-y)). \label{eq:4}
\end{align} 
Eq.~\eqref{eq:2} is equivalent to Eq.~\eqref{eq:PY_wf} and Eq.~\eqref{eq:4} is equivalent to Eq.~\eqref{eq:wf}. 
This argument can be easily extended to any number of particles and in any dimension and thus it becomes obvious 
that any $\Psi_{\bm n}({\bf r}_1,\ldots,{\bf r}_{A})$ 
does not depends on the CoM ${\bf R}_{\rm CoM}$ coordinate for any ${\bf n}$.
%%%%%%%%%%%%%%%%%%%%%%%%%%%%%%%%%%%%%%%%%%%%%%%%%%%%%%%%%%%%%%%%%%%
                                                     
In the ground state ${\bm n}={\bm 0}=(0,0,0).$
In the expansion in Eq.~\eqref{eq:PY_wf} there is only one state with CoM ${\bf P}\equiv {\bm 0}=(0,0,0)$. One can easily show that 
\begin{align}
& E_{\bf n} =\langle \Psi_{\bf n}| \hat{\rm H}-\hat{\rm K}_{\rm CoM}|\Psi_{\bf n}\rangle \equiv \langle \Psi_{\bf n}| \hat{\rm H}|\Psi_{\bf n}\rangle,\label{eq:En}\\
%&\langle \Psi_{\bf n}| \hat{\rm T}_{\rm CoM}|\Psi_{\bf n}\rangle \equiv 0,  \quad \forall \,{\bf n}, \label{eq:CoM}\\
&        \Psi_{\bm n}({\bf r}_1,\ldots,{\bf r}_{A})\neq \Psi_{\bm 0} ( {\bf r}_1,\ldots,{\bf r}_{A} ),  \, {\rm if} \, {\bf n} \neq {\bf 0}, \nonumber \\
&         E_{\bm n}\neq E_{\bm m}, \quad {\rm if} \quad {\bm n}\neq {\bm m}, \,{\rm and } \,\,\,E_{\bm 0} < E_{\bm n},\quad {\rm if} \quad {\bm n}\neq {\bm 0},\\
&\langle \psi |\hat{\rm H}-\hat{\rm K}_{\rm CoM}|\psi\rangle = \sumint_{\bm n}|C_{\bm n}|^2 E_{\bm n} \, {\bm >} \,  
{\bm E}_{\bm 0}, \, \sumint_{\bm n}|C_{\bm n}|^2=1, \label{eq:energy} \\
& \langle \Psi_{\bm 0} | \hat{\rm K}_{\rm CoM}|\Psi_{\bm 0}\rangle\, 
{\bm \equiv}\, 0,  \langle \Psi _{\bm 0}|\hat{\bf P}_{\rm CoM}|\Psi_{\bm 0}\rangle \, {\bm \equiv} \, {\bm 0}, \label{eq:P&Y} \\
&E_{gs}= E_{\bm 0}=\langle \Psi_{\bm 0}|\hat{\rm H}|\Psi_{\bm 0}\rangle\, {\bm <} \, \langle \psi |\hat{\rm H}-\hat{\rm K}_{\rm CoM}|\psi\rangle . \label{eq:egs}
\end{align}
Eqs.~(\ref{eq:En},  \ref{eq:P&Y}, \ref{eq:egs}) might appear to the reader unexpected. 
The reason why the above equations are correct is quite simple. In Eq.~\eqref{eq:wf}  the integral is performed 
over the CoM coordinate and as result the resulting integral $ C_{\bm n} \Psi_{\bm n}({\bf r}_1,\ldots,{\bf r}_{A})$ does not depend on the CoM coordinate 
${\bf R}$, as one can easily change the integration variable from ${\bf a}$ to ${\bf R}-{\bf a}$, which leaves the value of the integral unchanged. 
The CoM projected many-body wave functions  $\Psi_{\bf n}$ depend only on the $3(A-1)$ intrinsic relative (Jacobi) coordinates. 
Thus $\Psi_{\bm n}({\bf r}_1,\ldots,{\bf r}_{A})$ are the CoM projected intrinsic many-body wave functions for any CoM momentum ${\bf P}_{\bm n}$.  

Eq.~\eqref{eq:egs} is thus our most important formal result, which even though it is  so easy to obtain, it has never been discussed 
in literature as far as we aware. This equation explicitly demonstrate that the usual 
prescriptions used so far in literature to evaluate the CoM correction within mean field models with Eq.~\eqref{eq:t_com} 
and the corresponding corrected ground state energy 
using Eqs.~(\ref{eq:KE0}, \ref{eq:energy}) are  intrinsically inaccurate.

While Eqs.~(\ref{eq:KE0}, \ref{eq:t_com}) appear to provide an accurate evaluation of the overall size of the CoM kinetic energy fluctuations,
the expected ``intrinsic corrected mean field energy'' given by Eq.~\eqref{eq:energy} is contaminated 
by significant contributions from intrinsic states with corresponding total CoM momentum ${\bf P}_{\bm n} \nequiv 0$
 \begin{align}
&E_{\rm CoM}\,  \stackrel{\bf ???}{= } \,  -\langle \psi |\hat{\rm K}_{\rm CoM}|\psi\rangle = -\sumint_{\bm n}|C_{\bm n}|^2 \frac{ |{\bf P}_{\bm n}|^2}{2Am},\label{eq:t_com}\\
&|C_{\bm n}|^2 = \int \!\!\frac{d^3{\bf a}}{V} \exp  \left( i\frac{{\bf P}_{\bm n}\cdot{\bf a}}{\hbar}\right)  \langle \psi ( {\bf a} ) | \psi( {\bf 0} )\rangle \label{eq:P2} \\
&\quad \quad  \approx \frac{  (\sqrt{2 \pi}\sigma_{\mathcal{O} })^3 }{V} \exp\left ( -\frac{ |{\bf P}_{\bm n}|^2\sigma_{\mathcal{O} }^2}{2\hbar^2}\right )\ll 1,\label{eq:PP}\\
&\langle \psi |\psi \rangle = \sumint_{\bf n}|C_{\bf n}|^2=1,
\end{align}
see Fig.~\ref{fig:one} for more details. It comes as no surprise that the weights $|C_{\bm n}|^2$ of the projected many-body wave functions 
in the unprojected  mean field wave many-body wave function are indeed very small, but that means the the corresponding distribution is very wide. 
The widely used expression for the estimate of the CoM energy correction Eq.~\eqref{eq:KE0} would be accurate only if this distribution is 
extremely narrow, which in case of nuclei never seems to the case, see discussion below, Fig.~\ref{fig:ecom}, and Table~\ref{table:initial}, 
as the disagreement between $E_{\rm CoM}$ defined in Eq.~\eqref{eq:KE0} and Eq.~\eqref{eq:com} below increase with mass number $A$, 
which for nuclei with $A> 90$ leads to PY ground state energies lower by at least 1 MeV than the CoM corrected with Eq.~\eqref{eq:KE0} energies. 

The reason why $\langle \psi |\hat{\rm H}-\hat{\rm T}_{\rm CoM}|\psi\rangle$ leads to an inaccurate estimate 
of $E_{gs}$ is rather simple. In order to evaluate the ground state wave function $|\psi\rangle$ one perturbs the intrinsic Hamiltonian 
$\hat{\rm H}-\hat{\rm T}_{\rm CoM}$ by adding $\hat{\rm T}_{\rm CoM}$, while the intrinsic ground state  energy is determined by the
intrinsic translational invariant many-body wave function $|\Psi_{\bm 0}\rangle {\bm \neq} |\psi\rangle$. While the ``perturbation'' of the intrinsic Hamiltonian 
$\hat{\rm H}_{int}= \hat{\rm H}-\hat{\rm T}_{\rm CoM}$ by $\hat{\rm T}_{\rm CoM}$ is small compared to $\hat{\rm H}$, it is not accurate enough. In order to achieve a physically  
accurate estimate of $E_{gs}$ at least  second order perturbation contributions are required, which are negative for the ground state, 
as inequality in Eq.~\eqref{eq:egs} also confirms.

The PY approach is a different approach from simply ``trying to fix'' the lack of  translational symmetry 
with Eq.~\eqref{eq:KE0}, but never really restoring it.
The role of nucleons interaction is not encoded in the widely used CoM energy correction Eq.~\eqref{eq:KE0}, and
this correction is also contaminated by contributions from excited states, as Eqs.~(\ref{eq:energy}, \ref{eq:egs}) also clearly demonstrate. 
The mechanism of translational symmetry restoration is identical to that of the complete delocalization of 
the  electron wave  functions into the Bloch waves in conductors~\cite{Bloch:1929,Mermin:1976}. 
While the CoM motion in $\psi({\bf r}_1,\ldots,{\bf r}_A)$ is strongly localized, the wave functions 
$\Psi_{\bm n}({\bf r}_1,\ldots,{\bf r}_A)e^{  i {\bf P}_{\bm n}\cdot {\bf R}/\hbar }$ describe fully delocalized nuclear states.
The CoM spatial fluctuations volume is $ (\sqrt{2 \pi}\sigma_{\mathcal{O} })^3 = 3.86$ fm$^3$ and 0.17 fm$^3$ for  $^{16}$O and $^{208}$Pb, 
 corresponding to cubes of side 1.57 fm and 0.55 fm for $^{16}$O and $^{208}$Pb respectively, see Table~\ref{table:initial}. Using Eq.~\eqref{eq:PP} to estimate 
$E_{\rm CoM}$ in Eq.~\eqref{eq:t_com} leads, as expected, to values in close agreement with those reported in Table~\ref{table:initial}.
According to either our results, or using estimates of the CoM energy correction evaluated with Eq.~\eqref{eq:KE0} used in Ref.~\cite{Grams:2023},
we obtain that in the above equations Eqs.~(\ref{eq:energy}, \ref{eq:t_com}) $|C_{\bm n}|^2$ peak at a
CoM momentum $|{\bf P}_{\bm n}| =$ 315 MeV/c for $^{16}$O and 895 MeV/c for $^{208}$Pb, which 
points to a large  CoM momentum distribution in the nuclear mean field many-body wave functions. 
A wider CoM momentum distribution corresponds to a narrower spatial CoM distribution and using the approximate Eq.~\eqref{eq:PP} 
and the values reported in Table~\ref{table:initial} one can show that expected average of CoM kinetic energy, the negative of $E_{\rm CoM}$ defined in Eq.~\eqref{eq:KE0},
\begin{align}
E_{CoM} = \langle \psi |\hat{\rm K}_{\rm CoM}|\psi\rangle =\sumint_{\bm n} |C_{\bm n}|^2  \frac{ {\bf P}^2_{\bf n} }{2Am} \approx \frac{3\hbar^2}{2mA\sigma_\mathcal{O} ^2} \label{eq:MF11}
\end{align}
scales also  approximately as $A^{-1/6}$, as the exact PY CoM energy correction $E_{\rm CoM}$, see Eq.~\eqref{eq:com}, 
reported in Fig.~\ref{fig:ecom} and Table~\ref{table:initial}.
This result is based only the validity of the Gaussian overlap approximation (GOA) for the 
overlap kernels $\langle \psi({\bf a})|\psi({\bf 0})\rangle  \approx \exp[ - {\bf a}^2/(2 \sigma_\mathcal{O}^2 )  ]$, 
which appears to be satisfied essentially perfectly, see Fig.~\ref{fig:one}, a results which has been noticed by other authors in the past as well, see 
Ref.~\cite{Ring:2004} and many earlier references going back to \textcite{Griffin:1957}.

We ascribe the emergence of the GOA for the overlap kernels to the fact the EDF satisfies the generalization of the Galilean invariance 
under local gauge transformations~\cite{Engel:1975,Bender:2003,Bulgac:2018,Bulgac:2007}.  These transformations are valid even 
when the collective flow is not uniform, as in hydrodynamics ~\cite{LL2:1951,LL6:1959}. In such EDFs  
the CoM contributions to the EDF are clearly separated from the contributions due to intrinsic many-particle motion, 
as they are in the case of a many-body Hamiltonian, where 
\begin{align}
\hat{\rm H} = \hat{\rm K}_{\rm CoM} + \hat{\rm K}_{int}+\hat{\rm V}, \quad \hat{K} = \hat{\rm K}_{\rm CoM} + \hat{\rm K}_{int},
\end{align}
where $\hat{\rm K}_{int}$ depends only on relative momenta between particles. While in a mean filed many-body wave function there is no separation 
between the system CoM coordinate and the intrinsic coordinates, while the ground state energy is minimized so are the CoM fluctuations minimized
to the maximum possible extent, as they are ``entangled'' with the intrinsic coordinates. Since the 3D CoM coordinates are ``free'' the 
projected CoM wave function  becomes a Gaussian in order to minimize the uncertainty between the 3D CoM momenta and coordinates. 
While this is not an strict argument it can be accepted \emph{a posteori} as a justification of the ``Gaussianity'' of the projected 
CoM momentum distribution, see Eq.~\eqref{eq:PP}, which likely it cannot otherwise be mathematically proven.

The correct CoM energy correction to the binding energy is not given by either Eqs.~({\ref{eq:KE0}, \ref{eq:t_com}), but by 
\begin{align}
E_{\rm CoM}  =  \langle \Psi_{\bm 0}| \hat{\rm H} | \Psi_{\bm 0} \rangle- 
\langle\psi |\hat{\rm H}|\psi \rangle= E_{gs}-E_{MF}\leq 0, 
\label{eq:com}
\end{align}
which is how $E_{\rm CoM}$ was evaluated in this work and reported in Fig.~\ref{fig:ecom} and in Table~\ref{table:initial}. 
Eq.~\eqref{eq:MF11} represents an alternative to Eq.~\eqref{eq:KE0} to evaluate $E_{\rm CoM}$, which most likely  
less computational intensive that the direct evaluation of Eq.~\eqref{eq:KE0} used in practice~\cite{Da-Costa:2024}.

\begin{figure}
\includegraphics[width=0.9\columnwidth]{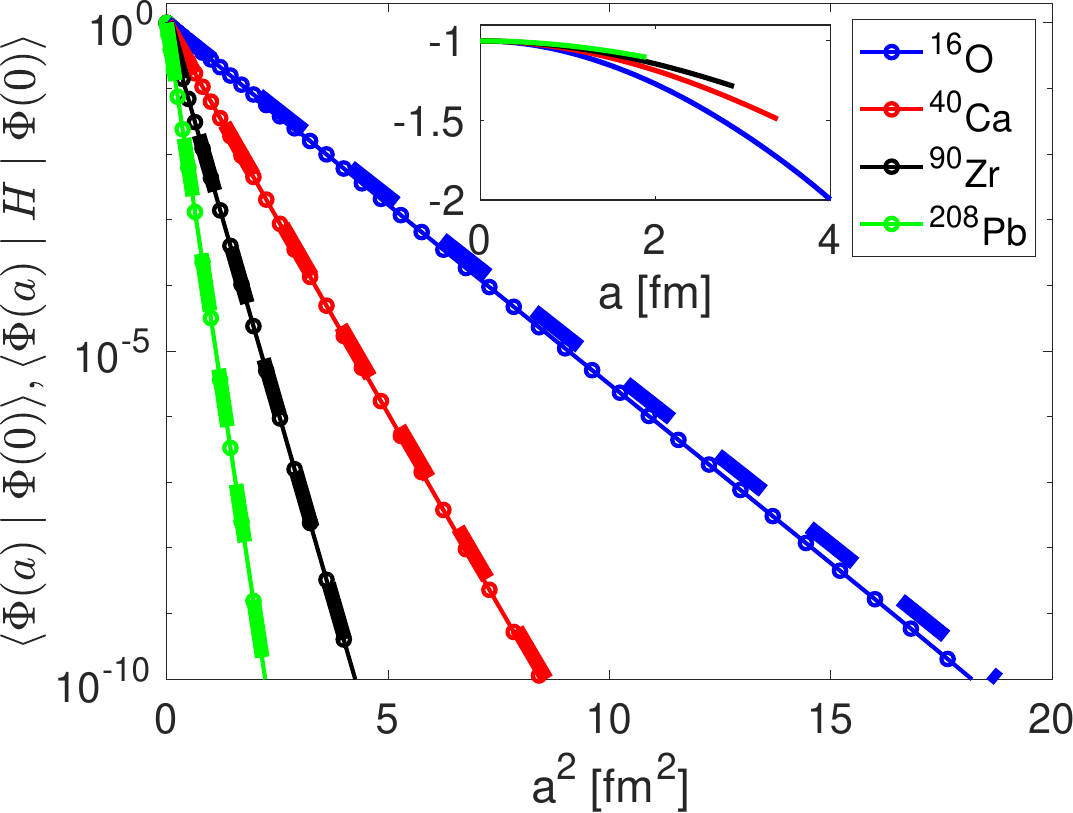}
\caption{ \label{fig:one}   The many-body wave functions overlap 
$  \langle \psi({\bf a})|\psi({\bf 0})\rangle  \approx \exp[ - {\bf a}^2/(2 \sigma_\mathcal{O}^2 )  ]$ solid lines and with dashed lines 
for the Hamiltonian overlap $ \langle \psi ( {\bf a} ) | H | \psi({\bf 0}) )\rangle$ normalized to 
$\langle \psi ( {\bf 0} ) | H | \psi( {\bf 0} )\rangle$. 
${\cal H}({\bf a},{\bf 0})$, see Eq.~\eqref{eq:gm}, normalized to the absolute value $|{\cal H}({\bf 0},{\bf 0})|$ 
is shown in the inset with lines of the same color for each nucleus.
Since all modern EDF satisfy a generalization of the Galilean invariance under local gauge 
transformations~\cite{Engel:1975,Bender:2003,Bulgac:2018,Bulgac:2007} and since $[\hat{\bf R},\hat{\bf P}_{\rm CoM}]=i\hbar$ it is 
not surprising that CoM fluctuations are Gaussian in character, see  Eq.~\eqref{eq:PP}. }
\end{figure}

The reader will easily recognize that the PY prescription 
is identical in spirit with the projection to zero total angular momentum
 in the case of an even-even deformed nucleus~\cite{Yoccoz:1957,Yoccoz:1957a,Ring:2004},
 \begin{align}
\hat{P}^0_{00} |\psi\rangle = \int \!\!  \frac{d\Omega}{8\pi^2} \,D^0_{00}(\Omega)\hat{R}(\Omega) |\psi\rangle ,
\end{align} 
and also similar to the projection to fixed particle number when pairing correlations are present.
In this manner all main mean field broken symmetries, translational, rotation, and gauge, 
are restored using exactly  the same method, which lead to many-body wave functions invariant under all these symmetries.

\begin{table}[h]
\centering
\begin{tabular}{ccccccccc}
\hline \hline
Nucl. & $E_\text{CoM}$ & $K_\text{CoM}$ & $r_\text{mat}$ & $\Delta r_\text{mat}$ &
$r_\text{ch}$ & $\Delta r_\text{ch}$ & $\sigma_\mathcal{O}$ & $\Delta \sigma$  \\ \hline
$^{16}$O & -10.05 & -11.20 & 2.66 & 0.058 & 2.67 & 0.058 &  0.626 & 0.010 \\
$^{40}$Ca & -9.20 & -9.62   & 3.38 & 0.020 & 3.40 & 0.021 & 0.429 &  0.003 \\
$^{48}$Ca & -9.03 & -9.68 & 3.53 & 0.015 & 3.44 & 0.014 & 0.392 & 0.002 \\
$^{90}$Zr &  -8.46 & -8.43   & 4.26 & 0.008 & 4.22 & 0.007 & 0.305 & 0.0009  \\
$^{100}$Sn & -8.31 & -8.31 & 4.39 & 0.006 & 4.43 & 0.006 & 0.290 & 0.0008 \\
$^{132}$Sn & -7.78 & -8.00 & 4.82 & 0.005 & 4.69 & 0.004 & 0.263 & 0.0005 \\
$^{208}$Pb&-7.21 & -7.26   & 5.57 & 0.003 & 5.49 & 0.003 & 0.220 & 0.0003 \\
\hline \hline
\end{tabular}
\caption{ $E_\text{CoM}$ and $K_\text{CoM}$ are defined in Eqs.~\eqref{eq:com} and \eqref{eq:kin},
$r_\text{mat}, \Delta r_\text{mat}, r_\text{ch}, \Delta r_\text{ch}$ are the matter and charge radii  
and their corresponding changes  after taking into account CoM fluctuations. Our results show an approximate 
$E_{\rm CoM}\approx (-17.3\pm 0.7)A^{-1/6}$ MeV and $K_{\rm CoM}\approx  (-17.93\pm  0.26)A^{-1/6}$   behavior.
The nucleon charge form factors have not been accounted for in this case. 
While evaluating charge densities the nucleon charge form factors are included as corrections using the $F^A_2$ 
convolution approximation~\cite{Frankfurt:1988,Segarra:2021,Denniston:2024}. 
The last two columns list the Gaussian width of 
the norm ${\sigma}_\mathcal{O}$ and the amount by which the Gaussian width of the 
Hamiltonian overlap is larger than Gaussian width of the norm overlap, $\Delta\sigma$.
}
\label{table:initial}
\end{table}

Our findings suggest also a potentially improved configuration interaction or equivalently a no-core shell model basis 
set of many-body wave functions including all $\Psi_{\bm n}({\bf r}_1,\ldots,{\bf r}_A)$, 
as Eq.~\eqref{eq:Psi2} is exactly fulfilled for the entire set.  With a proper choice of the size of this 
many-body wave functions set one can achieve any desired accuracy in principle.  

\section{Implementation of the translational symmetry restoration method} \label{sec:III}

A simpler formula for $E_{gs}$ also exists~\cite{Peierls:1957}
\begin{align}
E_{gs} = \frac{ \langle \Psi _{\bm 0}| \hat{\rm H} | \psi\rangle }{ \langle \Psi_{\bm 0} | \psi \rangle } = 
\frac{ \int d^3{\bf a} \langle \psi ({\bf a}) | \hat{\rm H} | \psi({\bf 0} )\rangle }{ \int d^3 {\bf a} \langle \psi ({\bf a}) | \psi ({\bf 0})\rangle }, \label{eq:GCM}
\end{align}
where one recognizes that this expression for $E_{gs}$ is a GCM
estimate~\cite{Griffin:1957,Ring:2004}.  
The many-body wave function in Eq.~\eqref{eq:wf} $\Psi_{\bm 0}({\bf r}_1,\ldots,{\bf r}_A)$ 
suggested by \textcite{Peierls:1957} in  1957 is exactly of the GCM form 
advocated a few months later in the same year by \textcite{Griffin:1957}. 
Using the overlap expressions translated into coordinate representation from Appendix E in  Ring and Schuck's 
monograph \cite{Ring:2004} for $\langle \psi ({\bf a}) | H | \psi({\bf 0} )\rangle$ we obtain,
\begin{align} 
&\!\!\!\!\! \langle \psi ( {\bf a} ) | \hat{\rm H} | \psi( {\bf 0} )\rangle = \langle \psi ( {\bf a} ) | \psi ( {\bf 0} )\rangle {\cal H}( {\bf a},{\bf 0} ), \label{eq:Ham}\\
&\!\!\!\!\!{\cal H}( {\bf a},{\bf 0} ) =   {Tr} ( \hat{\rm K}n^{{\bf a},{\bf 0}} )+
\frac{1}{2} {Tr_1} {Tr_2} ( n^{{\bf a},{\bf 0}} \hat{\rm V}_2 n^{{\bf a},{\bf 0}} ) + ...,  \label{eq:gm}\\
&n^{{\bf a0}}({\bf r},\sigma|{\bf r}',\sigma')= 
\frac{\langle \psi({\bf a})|\psi^\dagger({\bf r}',\sigma')\psi({\bf r},\sigma)|\psi({\bf 0})\rangle}
{\langle \psi ({\bf a}) | \psi ({\bf 0})\rangle},\label{eq:rho}
\end{align}
where $\hat{\rm K}$ stands for the one-body part of $\hat{\rm H}$ (typically the kinetic energy and a possible external field or a shape constraint), 
$\hat{\rm V}_2$ stands for the 2-body interaction, $\psi^\dagger, \,\psi$ are creation and annihilation field operators, $Tr$ stands for the trace,  
and ellipses are for terms arising from anomalous densities when present, three-particle interactions, etc.  
and ${\bf r},\sigma,\, {\bf r}',\sigma'$ stand for spatial, spin, 
and isospin coordinates.  The PY CoM energy correction Eq.~\eqref{eq:com}
to the nuclear binding energy points to the non-vanishing role played by the interactions between 
particles, even if the Hamiltonian $\hat{\rm H}$ is translational invariant, which is unaccounted for in Eq.~\eqref{eq:KE0}. 
This particular aspect was never discussed in literature, and was likely only mutely acknowledged when various authors 
used  an effective coordinate nucleon mass while evaluating $E_{\rm CoM}$. Strictly speaking Eq.~\eqref{eq:KE0} is an exact 
evaluation of the CoM kinetic energy only if $[\hat{\rm H},\hat{\rm K}_{\rm CoM}]\equiv 0$ and the CoM motion separates 
exactly in $\psi$, which is never the case in a mean field approximation.

The norm overlap $\langle \psi ({\bf a}) | \psi ({\bf 0})\rangle$ is given by determinant of the overlap matrix $M$ 
of the single-particle wave functions
\begin{align}
&M_{kl}({\bf a}) = \langle \psi_k({\bf a})|\psi_l({\bf 0)}\rangle, \, \langle \psi ({\bf a}) | \psi ({\bf 0})\rangle=\text{Det} M({\bf a}). \label{eq:wfs}
\end{align}
The one-body densities needed  to evaluate the binding energy in Eq.~\eqref{eq:GCM} 
are evaluated using the one-body density matrix 
\begin{align}
&n^{\bf a0}({\bf r},\sigma|{\bf r}',\sigma')= \frac{\langle \psi({\bf a})| \psi^\dagger({{\bf r}',\sigma'})\psi({\bf r},\sigma)
|\psi({\bf 0})\rangle}{\langle \psi ({\bf a}) | \psi ({\bf 0})\rangle} \nonumber \\
& \quad \quad \quad \quad \quad \quad 
=\sum_{kl} \psi_k({\bf 0}|{\bf r},\sigma) \rho^{\bf a0}_{kl}\psi_l^*({\bf a}|{\bf r}',\sigma') \label{eq:n}, \\
&\rho^{\bf a0}_{kl}=\frac{\langle \psi({\bf a})|c^\dagger_lc_k
|\psi({\bf 0})\rangle}{\langle \psi ({\bf a}) | \psi ({\bf 0})\rangle}=M^{-1}_{kl}({\bf a}), \label{eq:over}
\end{align}
\begin{align}
& \sumint_{{\bf r}',\sigma' }n^{\bf a0}({\bf r},\sigma|{\bf r}',\sigma')n^{\bf a0}({\bf r}',\sigma'|{\bf r}'',\sigma'')=\nonumber \\
& \quad \quad \quad  \quad \quad \quad n^{\bf a0}({\bf r},\sigma|{\bf r}'',\sigma''), \label{eq:proj}\\
& \sumint_{\bf r}n^{\bf a0}({\bf r},\sigma,{\bf r},\sigma) = A \quad (\text{total particle number}),\\
& \psi_k({\bf 0}|{\bf r},\sigma) = \langle 0| \psi({\bf r},\sigma) c_k^\dagger|0\rangle, \, |\psi_k({\bf 0})\rangle = c_k^\dagger|0\rangle,  \\
& \psi^\dagger({\bf r}',\sigma') = \sum_l \phi_l^*({\bf r}',\sigma')c_l^\dagger,\\
& \psi({\bf r},\sigma) = \sum_k\phi_k({\bf r},\sigma)c_k, \, |\psi({\bf 0)}\rangle = \prod_mc_m^\dagger|0\rangle, 
\end{align}
and $\psi^\dagger({\bf r}',\sigma')$ and $\psi({\bf r},\sigma)$ are creation and annihilation field operators and $|0\rangle$
is the vacuum state and $\phi_{k,l}({\bf r},\sigma)$ are single-particle wave functions.  
$n^{\bf a0}({\bf r},\sigma,{\bf r}',\sigma')$ is a projector and was introduced 
by \textcite{Lowdin:1955a}  and it can be used to evaluate the Hamiltonian overlap 
$\langle \psi ( {\bf a} ) | \hat{\rm H} | \psi( {\bf 0} )\rangle$ for any type of interactions between fermions.
In the case of a Skyrme-like EDF  the one-body (diagonal) number densities in coordinate representation
$\sum_\sigma n^{\bf a0}({\bf r},\sigma|{\bf r},\sigma)$ and the other needed kinetic, current, and spin-orbit densities
enter in the EDF exactly as in the case of familiar Skyrme-like Hartree-Fock expression. 
Skyrme-like EDF are functionals defined as functions of  nucleon number 
densities and their spatial derivatives, nucleon currents, spin-densities and corresponding currents, nucleon kinetic energy densities, 
and anomalous densities. They are typically required to satisfy a generalization of the Galilean invariance under local gauge transformations~\cite{Engel:1975,Bender:2003,Bulgac:2018,Bulgac:2007}.
This is the approach adopted by  \textcite{Kohn:1965} to generate a Hartree-like EDF in 
terms of  number densities and uniformly used  nowadays
for electronic systems~\cite{Gross:2006,Gross:2012}, and also in agreement with current approaches 
for finite systems~\cite{Sheikh:2021}, when symmetries are restored. The case of the Unitary Fermi 
Gas~\cite{Bulgac:2007,Bulgac:2011a,Bulgac:2013a,Bulgac:2019}is a case in point, 
where fractional powers of number densities appear as required by dimensional arguments. 
This is in agreement with \textcite{Hohenberg:1964} theorem that the exact many-body wave functions 
are (underivable) functionals of the number density alone. 

In the case of closed-shell systems with total angular momentum zero the evaluation 
of the overlap matrix elements $\langle \psi ( {\bf a} ) | H | \psi( {\bf 0} )\rangle$ and 
$\langle \psi ( {\bf a} ) | \psi( {\bf 0} )\rangle$ is technically much simpler, 
\begin{align}
& \!\!\!\!\! \int \!\!\!d^3{\bf a} \langle \psi ( {\bf a} ) | \hat{\rm H} | \psi( {\bf 0} )\rangle 
\equiv 4\pi \int_0^\infty \!\!\!da\, a^2 \langle \psi ( { a} ) | \hat{\rm H} | \psi( { 0} )\rangle,\\
&\!\!\!\!\! \int \!\!\!d^3{\bf a} \langle \psi ( {\bf a} ) | \psi( {\bf 0} )\rangle 
\equiv 4\pi \int_0^\infty \!\!\!da\, a^2 \langle \psi ( { a} ) | \psi( { 0} )\rangle,
\end{align}
when the shift in ${\bf a}$ can be evaluated only in one spatial direction, e.g.  ${\bf a} = (0,0,a)$.
Both norm and Hamiltonian overlaps 
appear to be essentially Gaussian over a wide range of displacements ${\bf a}$, see Fig.~\ref{fig:one}. 

In nuclei of arbitrary shapes in a mean field framework one can use only as little as 3 values for ${\bf a}$ 
in carefully chosen principal directions in 3D  in order to evaluate the corresponding Gaussian widths 
and speed-up considerably the evaluation of the $E_{\rm CoM}$ correction, as an alternative to Eqs.~\eqref{eq:KE0}.

Apart from the $E_{\rm CoM}$ correction,  it is particularly informative to compare the 
change in the total kinetic energy, and the change  of mass and charge RMS radii after CoM projection.  
The changes in the density distributions after CoM projection are relatively small.
\begin{align}
&K_{\rm CoM} = \langle \Psi _{\bm 0} |\hat{\rm K} | \Psi_{\bm 0}\rangle  -\langle \psi|\hat{\rm K}|\psi\rangle, \label{eq:kin}\\
&\hat{n}({\bf r}) = \sum_\sigma \psi^\dagger({\bf r},\sigma)\psi({\bf r},\sigma) ,\\
&\Delta n({\bf r}) =  \langle \Psi _{\bm 0}| \hat{n}({\bf r})| \Psi_{\bm 0}\rangle 
- \langle\psi |\hat{n}({\bf r})|\psi \rangle , \label{eq:raddif}\\
&(\Delta r_{ch,mat})^2= \int \!\! d^3{\bf r} \,{\bf r}^2\Delta n({\bf r}), \,
 \int \!\!d^3{\bf r}\Delta n({\bf r}) \equiv 0, \label{eq:rad}
\end{align}
where $\hat{n}({\bf r})$ stands for either the total number density matter distribution 
or  the point proton or neutron number distribution respectively. 
In Table I we summarize our findings for  magic 
nuclei $^{16}$O, $^{40}$Ca,  $^{48}$Ca, $^{90}$Zr, $^{100}$Sn, $^{132}$Sn, and $^{208}$Pb, 
where we display $E_{\rm CoM}$~\eqref{eq:com}, $T_{\rm CoM}$~\eqref{eq:kin}, 
and the matter and charge distribution radii and their change before and after CoM projection as 
defined in Eq.~\eqref{eq:rad}. We also list the Gaussian widths of the many-body wave function 
and Hamiltonian overlaps, which are both clearly Gaussian in character, see Fig.~\ref{fig:one}, as a function of the offset $a$.
All calculations were performed with the EDF SeaLL1~\cite{Bulgac:2018}, which provides one of the best 
descriptions within DFT, extended to include pairing correlations, of a rather large range of nuclear properties across the entire 
nuclear chart, an rms accuracy for the total binding energy  within mean field 
reached with only seven parameters: equilibrium energy and density of homogenous nuclear matter, 
symmetry energy and its density dependence, surface tension, and pairing and spin-orbit couplings, albeit without any 
symmetry and zero-point fluctuations corrections.  The self-consistent calculations for  
these nuclei were performed in a cubic box large enough to fit each nucleus and with a lattice constant $l=1$ fm, 
a value which insures a good accuracy in both static and dynamic DFT simulations~\cite{Ryssens:2015,Shi:2020,Bulgac:2023}. 

In Refs.~\cite{Schmid:1990,Schmid:1991} the authors consider $^{16}$O with a different EDF and 
they observed differences between the projected and unprojected number densities 
for radii $r< 1.5$ fm or so, which are comparable to our results obtained for a lattice constant $l=1$ fm. These authors
were interested in the charge form factor at transferred momenta $q> 2$ fm$^{-1}$, where 
the charge form factors change at the level $10^{-4}$ with respect to the maximum, which is consistent 
with our results.  We could not compare our results for the energies with those 
of Ref.~\cite{Rodriguez-Guzman:2004a} for $^{16}$O and $^{40}$Ca however, for reasons explained above.

The DFT approach parallels what is done in the case of the  many nucleon Schr\"odinger  equation, 
where the parameters of the interparticle interactions are extracted from phenomenological
studies, based on a reduced knowledge of the energy-dependence and partial wave decomposition of the cross sections 
and some inferences from QCD, extrapolated to low momenta below 
$\Lambda_{QCD}=600\ldots 1000$ MeV (thus incomplete). In such 
studies, often denoted as ``\textit{ab initio}'', the number of parameters for the inferred two-nucleon and 
three-nucleon interactions is larger than  20 for the 2-body part 
alone~\cite{Somasundaram:2024,Tichai:2024} and the deviation of the binding energy of $^{208}$Pb 
is at the level of about  200 MeV~\cite{Hu:2024}, which is worse by about two orders
of magnitude than the accuracy of phenomenological DFT models~\cite{Bender:2003,Da-Costa:2024,Grams:2023,Bulgac:2018} or
the RMS of the Bethe-Weizs\"acker mass formula. 
Low energy nuclear microscopic approaches cannot yet access distances corresponding to 
nucleon-nucleon separations less than the sum of their radii, thus less than about  $\pi\hbar c/\Lambda_{QCD}={\cal O}(1)$ fm.
There are notable efforts however to extend the reach of {\it ab initio} approaches to large transferred momenta, 
but that approach is limited to very light nuclei so far~\cite{Tropiano:2022,Tropiano:2024}, 
but see also Refs~\cite{Frankfurt:1988,Segarra:2021,Denniston:2024}. In {\it ab initio} calculations the evaluation of 
nucleon densities, which are one-body operators by Nature, require adding rather convoluted and physically rather opaque
2- and 3-body contributions~\cite{Tropiano:2024}, derived from many nucleon wave functions, 
sometimes requiring multi-dimensional integrations over $3A$ coordinates, which limits the applicability of 
such methods to light or medium and nuclei~\cite{Lonardoni:2017}.  Accurate energy-density functionals are at the same time easier 
to implement over the entire nuclear mass table and have a much higher accuracy.

\section{The Road Ahead} \label{sec:IV}

The first notable thing shown here is the quite significant magnitude of the CoM energy 
correction to the binding energy of nuclei and its mass dependence. While we cannot really compare our results with results 
obtained by other authors, who have used basically different EDFs, it is notable that our $|E_{\rm CoM}|$ are consistently 
larger by roughly 1 MeV or more for medium and heavy mass nuclei than the rest of the results available in literature, 
which is in agreement with the message of Eq.~\eqref{eq:egs} and reproduced below.
The translational-invariant many-body wave functions provide a lower estimate for nuclear binding energy than mean field result 
obtained with Eq.~\eqref{eq:KE0} CoM energy correction, which so far was as a rule used in literature, with the exception of a 
few works for nuclei with $A\leq 40$ reviewed by \textcite{Schmid:2004} and the singular work known to us by 
\textcite{Dobaczewski:2009}.
These are the few works in literature using the \textcite{Peierls:1957} framework known to us, 
\begin{align}
E_{gs}=\langle \Psi_{\bm 0}|\hat{\rm H}|\Psi_{\bm 0}\rangle\, {\bm <} \, \langle \psi |\hat{\rm H}-\hat{\rm K}_{\rm CoM}|\psi\rangle .  \nonumber
\end{align}
Here $|\Psi_{\bm 0}\rangle$ and $\psi$ are the restored translational invariant  and the mean field many-body wave functions respectively.

The next somewhat unexpected  fact  is that particle interactions appear to play a secondary role  
in the CoM correction to the binding energy, since $E_\text{CoM}\approx T_\text{CoM}$, 
with the exception of the lighter nuclei. In  light nuclei is where the effects of finite range nucleon interactions
are enhanced due to a more pronounced role  played by the nuclear surface.  
These results were obtained with the SeaLL1 EDF~\cite{Bulgac:2018}, where the only seven parameters are 
determined by some general nuclear  properties well-known for many decades, 
reaching also one of the best accuracies for RMS of mass defects of even-even nuclei in literature in 
the absence of BMF corrections,  and even though SeaLL1 energy density functional was not even optimized. 
In most DFT studies the number of parameters needed 
is between 15 or more, when including BMF corrections, see Refs.~\cite{Bender:2003,Da-Costa:2024,Grams:2023}.

With foreseeable improvements DFT will likely provide an even higher  
increased accuracy, particularly  when relativistic corrections will be correctly accounted for as well. 
For example, the average nucleon kinetic energy  is lowered by $\approx 0.2$ MeV with relativistic 
corrections and thus a nucleus mass can be noticeably affected, see Table~\ref{table:tab2}.

All non-relativistic DFT implementations  assume that nuclear masses satisfy
\begin{align} 
&\frac{M_A({\rm exp}) }{N \,m_n+Z\, m_p^2 }\equiv 1, \label{eq:inertia00}
\end{align}
where $m_{n,p}, \, M_A({\rm exp})$ are the measured neutron, proton and nuclear masses respectively,
a statement which goes all the way back to \textcite{Peierls:1957}  and it is often related to 
Galilean invariance~\cite{Schmid:2004,Bender:2003,Dobaczewski:2009,Sheikh:2021}, and not 
merely to Galilean translational invariant wave functions. 
While the Galilean invariance is built in the Schr\"odinger equation and in non-relativistic DFT,
 any predicted nuclear inertia are in strong violation with Nature, since 
\begin{align} 
&\frac{M_A ({\rm DFT})\,c^2}{N \,m_n\,c^2+Z\, m_p^2\,c^2 }< 1, \label{eq:inertia11}\\
&\frac{ M_{^{208}Pb }( {\rm SeaLL1})}{N \,m_n+Z\, m_p^2 } =0.991624,
\end{align} 
thus a violation of Galilean invariance of $\approx 1$\%. Dobaczewski~\cite{Dobaczewski:2009,Sheikh:2021} in 
his Peierls  \& Yoccoz's implementation of translational-symmetry restoration obtains
\begin{align} 
&\frac{M_A ({\rm DFT})\,c^2}{N \,m_n\,c^2+Z\, m_p^2\,c^2 } = 1.311(14)\,A^{-0.060(3)}, \label{eq:inertiaDD}
\end{align}
with values $\approx 1.095$ for $A=20$  and $\approx 0.952$ for $A=208$, thus 
with significant consequences on many observables.

\begin{table}[h]
\centering
\begin{tabular}{ccccc}
\hline \hline
Nucl. & $K_\text{Non-Rel.}$ & $K_\text{Rel.}$ & $\Delta K$ & $\Delta K/A$  \\ \hline
$^{16}$O & 239.92 & 237.10 & -2.83 & -0.18 \\
$^{40}$Ca & 660.91 & 652.87 & -8.03 & -0.20 \\
$^{48}$Ca & 851.08 & 840.10 & -10.98 & -0.23 \\
$^{90}$Zr & 1629.33 & 1608.53 & -20.80 & -0.23 \\
$^{100}$Sn & 1839.93 & 1816.33 & -23.59 & -0.24 \\
$^{132}$Sn & 2472.26 & 2439.90 & -32.36 & -0.25 \\
$^{208}$Pb & 3892.72 & 3842.56 & -50.16 & -0.24 \\
\hline \hline
\end{tabular}
\caption{ \label{table:tab2}
Kinetic energy evaluated with the non-relativistic dispersion and with the relativistic dispersion, using the non-relativistic 
HF single-particle wave functions. }
\end{table}

Even the first order relativistic correction to the mean field result to the inertia
due to relativistic corrections $\Delta K = -50.16$ MeV for $^{208}$Pb, see Table~\ref{table:tab2} is exceeding any BMF corrections studied so far in literature.
As a results the real inertia of any initial colliding heavy-ions and the corresponding reaction fragments for example, or of the 
fission fragments in fission is not correctly reproduced in and TDDFT calculations. In particular the total kinetic energies of the 
fission fragments is $\approx 177.8 -0.3489 \,E_n \pm 0.2$ MeV for  induced fission of $^{239}$Pu, where $E_n$ is the incident energy of the neutron, 
and change when the incident energy varies in the interval 5 MeV decreases from $\approx 177.9\pm 0.2$ to $\approx 175.9\pm 0.2$ MeV~\cite{Madland:2006}.
The  relative errors are lower by about an order of magnitude than the relative errors 
due to  DFT nuclear masses described in nonrelativistic TDDFT simulations~\cite{Bulgac:2016,Bulgac:2019c,Bulgac:2020,Bulgac:2022b},  
which all modern theoretical approaches are expected to reproduce and none of them do so far.

The overwhelming majority modern non-relativistic EDF in use in both static and dynamic simulations 
satisfy the local Galilean invariance, as described by \textcite{Engel:1975} 
more than 50 years ago,  in which the relevant densities and the corresponding mean fields are classical in character~\cite{Bertsch:1975}.
A generalization to Lorentzian local transformations as in relativistic classical theory of fields and hydrodynamics~\cite{LL2:1951,LL6:1959} is clearly 
required. 

The Walecka type of relativistic mean field used quite widely in literature~\cite{Walecka:1974,Meng:2023} 
are not really relativistic DFT frameworks in spirit, as they depend of number densities of non-constituents.
A true nuclear EDF is defined in terms of various number densities of the constituent nucleons. One of the main reasons 
why Walecka model is so popular, apart from allowing to extend studies to higher nuclear densities, is that it produced an apparently natural 
explanation to the nature of nuclear spin-orbit coupling, as a result of the competing roles in binding due to scalar and vector mean fields. 
TOn the other hand, the strength of the spin-orbit interaction appears to be explained microscopically lately by the role of 
the three-nucleon interactions~\cite{Hu:2022,Ding:2026}. 
In Walecka models one introduces scalar and vector very massive mesons and  often non-local couplings to nucleon fields.  

Other potential non-constituents, such as the Dirac sea states and pions, are neglected in Walecka type of EDF, and without a solid argument. 
The existence of real pions inside nuclei is totally unrealistic~\cite{Bulgac:1997}, as their localization energy in finite nuclei is very high. 
Pions have much smaller masses than the scalar and vector mesons typically considered, 
their presence would be equivalent to a real pseudo scalar field (aka ``pion condensation''), 
which require the mixing of at least two different Slater 
determinants with different spatial parities. Low energy nuclear states with opposite 
parities and identical total angular momenta but with different parities are separated by large energies of the order of several tens of MeVs, 
which are much smaller than the pion masses and their existence is thus not due to the existence of a real ``pion condensate'' in nuclei. 
The shell corrections due to the corresponding much stronger spin-orbit splittings 
experienced by the Dirac sea nucleon states are expected to be significant, since the strength of the spin-orbit interaction for Dirac sea nucleon states is 
increased by more than an order in magnitude compared to phenomenological nuclear spin-orbit strength~\cite{Bohr:1969}, 
for which so far there is  no observational support, such as the occurrence of magic nuclei~\cite{Mayer:1949,Haxel:1949}. 
Moreover, the effective nucleon mass in Walecka type of EDF is significantly reduced with respect to the bare nucleon mass, thus 
leading to a nuclear low single-particle energy level density. The observed single-particle level densities near the Fermi level in nuclei 
do not support small nucleon effective masses~\cite{Alex-Brown:1998,Kortelainen:2010,Kortelainen:2012,Bulgac:2018}.

The error arising form neglecting the relativistic corrections to nuclear masses are significantly greater than 
any other BMF corrections of the nuclear DFT. The introduction of relativistic corrections can be achieved with 
relatively little effort however. One has to remember the EDF 
has a classical character~\cite{Bertsch:1975}, since it is a function of various number densities, which are expectation values 
evaluated to the DFT wave functions of the real quantum operators $\psi^\dagger(\xi)\psi(\xi'), ...$, where $\xi=({\bf r},\sigma,\tau)$. 
In EDF the introduction of exchange number densities with ${\bf r}\neq {\bf r'}$ 
is a topic not fully resolved yet in theory. That is the reason why \textcite{Kohn:1965} introduced exchange and correlation energy
contributions to EDF on the same footing as a functional of the number density $n({\bf r})$ only, 
since a separation into pure exchange and pure correlations contributions 
to the total energy is very ambiguous in general beyond first order perturbation in two-body interaction, 
see Refs.~\cite{Karasiev:2014,Kas:2019,Bulgac:2023b} and other references therein.  
Following the consequence of the mathematical theorem of \textcite{Hohenberg:1964}'s stating that many-fermion wave functions are functionals 
of the number density $n({\bf r})$  alone, an EDF for a given quantum many fermion system is always constructed form 
various number densities. The mathematical procedure on how to construct a many-body wave function 
from the number density alone is however unknown, while the reverse process is trivial. 

 \begin{figure}
\includegraphics[width=0.9\columnwidth]{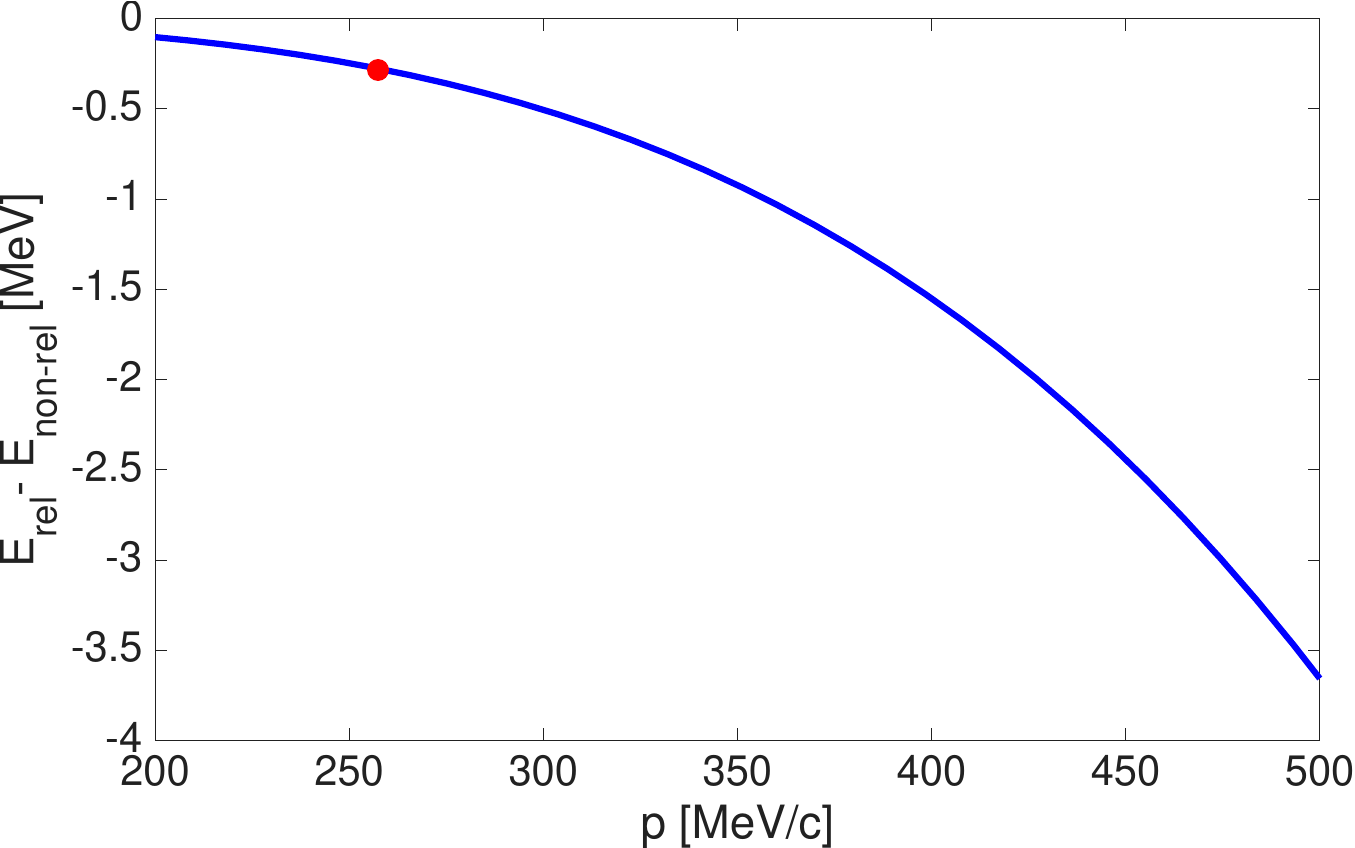}
\caption { \label{fig:E_kin}   The difference  between 
the relativistic and non-relativistic  expressions for the kinetic energy $E$ of a Fermi gas, see Eqs.~(\ref{eq:sq_exp}, \ref{eq:non-rel}, \ref{eq:delta_E} ). 
The red dot corresponds to the value of the Fermi momentum $p_F =257.4$ MeV/c in symmetric nuclear matter 
with equilibrium density $n=0.155$ fm$^{-3}$.  A momentum $p=500$ MeV/c would correspond to a projectile with the non-relativistic kinetic energy 
133 MeV/A impinging on a target at rest, an overestimate of the true kinetic energy $\approx 3\%$.}
\end{figure} 

One might be naturally tempted to limit CoM energy correction to  the first order relativistic correction to the total kinetic energy of a nucleus, 
the term $X^4$ in the equation below. In the relativistic Fermi gas approximation the first several terms of the kinetic energy per particle are 
\begin{align}
&\!\!\! \frac{E_{kin}}{A} \approx \frac{3mc^2}{10} \left ( X^2- \frac{5}{28}X^4 + \frac{5}{72}X^6-\frac{25}{704}X^8 +\ldots \right ),\label{eq:sq_exp}\\
&\!\!\! X =\frac{p_Fc}{mc^2} = 0.277, \quad \varepsilon_F = \frac{p_F^2}{2m} = 36.06 \, {\rm MeV},
\end{align} 
where the numerical values are evaluated at equilibrium symmetric nuclear matter, where $p_F = 257.4$ MeV/c, see Fig.~\ref{fig:E_kin}.
The infinite series in Eq.~\eqref{eq:sq_exp} is a conditionally 
converging  alternating infinite sum of the  function  $F(X)$
\begin{align}
&\frac{E_{kin}}{A} = \frac{3mc^2}{10} F(X),\\
& F(X)=\frac{10\int \!\!dX\, X^2(\sqrt{1+X^2}-1)}{3\int \!\! dX \, X^2}< X^2,  \label{eq:FX}
\end{align}
which can be derived using the binomial expansion of $\sqrt[\alpha]{1+X^2}-1$, and which  for any real number $0<\alpha\leq 1$ is 
a conditionally convergent infinite sum.   Since 
\begin{align}
\sqrt{1+X^2}-1 < X^2,
\end{align}
the relativistic kinetic energy of a Fermi gas is always strictly smaller than its non-relativistic counterpart for any size of the Fermi sphere, see Fig.~\ref{fig:E_kin}.
In case of a quantum operator  $\hat{\rm X}^2$ the energy spectrum of $\sqrt{1+\hat{\rm X}^2}-1$ will always be bounded from above by the
corresponding non-relativistic spectrum of $\hat{\rm X}^2$. The first relativistic correction to the kinetic energy in symmetric homogeneous nuclear matter in equilibrium 
per particle in Eq.~\eqref{eq:sq_exp} is clearly unexpectedly large and negative
\begin{align}
& \left . \frac{E_{kin}}{A} \right |_0 =  \frac{3\varepsilon_F}{5}= \frac{3mc^2}{10}  X^2 = 21.16 \, {\rm  MeV}, \label{eq:non-rel}\\
&\Delta_{1st } \left ( \frac{E_{kin}}{A}\right )  = -\frac{3mc^2}{10}\frac{5}{28}X^4 \approx - 0.284 \, {\rm MeV}, \label{eq:delta_E}
\end{align}
which for $^{208}$Pb would be approximately a correction 
of $\approx -59$ MeV to the total kinetic energy and consistent with the result in Table~\ref{table:tab2}. It makes no sense to absorb these corrections to the EDF 
as arising from interactions between nucleons. Including several terms of this expansion for the kinetic energy into EDF would make the full nuclear EDF  comparable 
in mathematical complexity with studied in the past EDFs, which included higher 
order corrections inspired by chiral effective field theory~\cite{Carlsson:2008,Carlsson:2010a}. 
The emerging effective mass depends in this case apart from powers of number 
density $n({\bf r})$  also on powers of kinetic energy density $\tau({\bf r})$, which has been considered before 
by other authors~\cite{Carlsson:2008}. One can always eschew using the explicit square root in EDF by including a controlled number of terms 
in a Taylor expansion of Eq.~\eqref{eq:FX}, see Eq.~\eqref{eq:sq_exp}.

It is always preferable to use an exact simple expression instead of a conditionally converging infinite series, 
which is what we propose below. Nature sometimes favors the presence of fractional powers of the 
number density in the energy of various systems, and the free Fermi gas and the Unitary Fermi Gas 
are prime examples~\cite{Bulgac:2007,Bulgac:2011a,Zwerger:2011}. As we have recently shown in 
Ref.~\cite{Kafker:2026}, in the case of SeaLL1 EDF there are no issues with performing particle projections 
for emerging fragments in the reaction the $^{48}$Ca+$^{208}$Pb as well as in the case of 
presence of pairing correlations~\cite{Bulgac:2024a}, unlike the issues discussed  in Refs.~\cite{Duguet:2009,Sheikh:2021}, where 
Hartree-Fock/Hartree-Fock-Bogoliubov (HF/HFB) approximations are conflated with DFT. In DFT the EDF is always 
expressed through various number densities, following \textcite{Kohn:1965} prescription, which is inline with the \cite{Hohenberg:1964} 
theorem that the many-body wave function is a functional of the number density alone. This thus implies that any two-body matrix element
$\langle \Phi|\psi^\dagger(\xi_1)\psi^\dagger(\xi_2)\psi(\xi_2')\psi(\xi_1')|\Phi\rangle$ is also a functional of  
$\langle \Phi|\psi^\dagger(\xi)\psi(\xi)|\Phi\rangle$, unlike in HF/HFB frameworks.
AB in collaboration with two other authors,  introduced the quantum fractional Fokker-Planck equation~\cite{Kusnezov:1999}, 
in order to accommodate anomalous diffusion in a Markov quantum process, such as L\'evy flights, which was a generalization of the quantum Fokker-Planck 
equation studied by the same three  authors in a range of papers. Fractional powers of various operators are rather common occurrences in literature nowadays. 

As in the case of restoring the local Galilean invariance~\cite{Engel:1975,Bender:2003,Bulgac:2018,Bulgac:2007} 
one has to replace the non-relativistic EDF the kinetic energy density with the corresponding relativistic expression,
by introducing the local collective momentum density ${\bf P}_{coll}({\bf r})$, 
the corresponding local single-particle intrinsic densities ${\bf p}_{int\, k} ({\bf r})$, and by separating the kinetic density 
into its collective and its intrinsic parts~\cite{Bulgac:2019c} 
\begin{align}
& {\bf P}_{coll}( {\bf r} ) = A\, {\bf p}_{coll}({\bf r})=\sum_k {\bf p}_k({\bf r}),  \label{eq:P_coll}\\
&{\bf p}_{k}({\bf r}) ={\rm Re} [ -i\hbar \phi_k({\bf r}){\vec \nabla} \phi_k^*({\bf r})],   \\
& {\bf p}_{int\, k} ( {\bf r} )= {\bf p}_k( {\bf r} ) - {\bf p}_{coll} ( {\bf r} ),\\
&  n_k({\bf r}) = |\phi_k({\bf r})|^2, \quad n({\bf r}) = \sum_k n_k({\bf r}), 
\end{align}
and by introducing their corresponding relativistic extensions 
\begin{align}
&\!\!\!\!\! K_{coll}({\bf r})  =  \sqrt{ 
    \left [  n({\bf r}) mc^2+{\cal E}_{int}({\bf r}) \right ]^2 + \left  |{\bf P}_{coll}({\bf r})c\right |^2  
                                                    }       \nonumber \\
& \quad \quad \quad \quad  - \bigl [n({\bf r})mc^2+{\cal E}_{int}({\bf r})\bigr] \label{eq:tcol} \\                                  
&\!\!\!\!\!  {\cal E}_{int} ({\bf r}) = \sum_k \sqrt{ [ n_k({\bf r}) \bigl (mc^2+\Upsilon ({\bf r}) \bigr )]^2 +|{\bf p}_{int\,k}({\bf r})c|^2 },  
%\nonumber\\
%&\!\!\! \quad \quad \quad \quad -\sum_kn_k({\bf r})\bigl (mc^2+\Upsilon({\bf r}) \bigr )  , 
\label{eq:tint} \\
&\!\!\! U (n({\bf r}), \tau({\bf r}),\ldots ) =   n({\bf r})\Upsilon({\bf r}),  \,
%&\!\!\! 
\tau({\bf r})= \sum_k | {\vec \nabla} \psi_k ({\bf r}) |^2,\\
%&\!\!\! {\cal E}_{int}({\bf r}) =K_{int}({\bf r}) +  U (n({\bf r}), \tau({\bf r}),\ldots ), \\ 
& \!\!\! E_{tot} = \int \!\! d^3{\bf r} \,{\cal E}({\bf r})=\int \!\! d^3{\bf r} \,\bigl [  K_{coll}({\bf r})  +{\cal E}_{int}({\bf r})\bigr ] , \label{eq:rDFT} 
\end{align}
where ${\cal E}({\bf r})$ is our suggested relativistic extension for a Skyrme-like EDF, 
which is locally Lorentzian invariant. All quantities in Eqs.~(\ref{eq:P_coll}-\ref{eq:rDFT}) are expectation values of quantum operators, 
and therefore they do not account for any quantum fluctuations, and have a classical  character~\cite{Bertsch:1975}.
For simplicity we do not display the spin and isospin variables in $\psi_k({\bf r})$. The  
$v_k$-components of the occupied quasiparticle wave functions 
should be used instead of $\psi_k({\bf r})$ if pairing correlations are present. 
Notice that the local Fermi ``sphere'' is at rest,  as $\sum_k {\bf p}_{int\, k}({\bf r}) \equiv 0.$ 
In the above equations the spin and isospin quantum numbers
($\sigma=\pm 1/2, \, \tau=\pm 1/2)$ of freedom are also  implicitly included in the index $k$.
$U (n({\bf r}), \tau({\bf r}),\ldots )$ is the intrinsic energy density due to the nucleon-nucleon interactions. 
Since ${\cal E}_{int}({\bf r})$ is defined the local reference frame, where 
\begin{align}
{\bf P}_{coll}( {\bf r} )=M_A{\bf V}_{coll}({\bf r}) \equiv 0,
\end{align} 
where $M_A$ is the nucleus inertia, see Eq.~\eqref{eq:PY_mass},  the collective momentum 
does not have to appear in the combination  $\hbar^2 \tau({\bf r})n({\bf r}) -[{\bf P}_{coll}({\bf r})]^2$, 
as in a non-relativistic Skyrme-like EDF~\cite{Engel:1975}. In particular, the ground state energy density of a relativistic 
Fermi gas or of a uniform relativistic Unitary Fermi gas is given exactly (up to the overall dimensionless 
Bertsch parameter) by  Eq.~\eqref{eq:tint} with both $n_k=1/V$ ($V$ being the volume) 
and ${\bf p}_{int\, k}$ coordinate independent. 

Dirac has considered such a form for the wave equation of a particle 
\begin{align}
{\text `}{\text `}\left \{ p_0 -\bigl ( m^2c^2 +p_1^2+p_2^2+p_3^2\bigr )^{\tfrac{1}{2}}\right \} \psi = 0,{\text '}{\text '}, \label{eq:Dirac}
\end{align}
{\it where the $p$'s are interpreted as operators ...}, but he concluded that the presence of the square root was unsatisfactory,
{\it because it is very unsymmetrical between $p_0$  and the other $p$'s'}, see Refs.~\cite{Dirac:1928,Dirac:1930,Dirac:1957}, 
and not because of the absence of Lorentz symmetry.  Notice that this relation, following Dirac's line of reasoning, {\it is very unsymmetrical between 
$p_0$  and the other $p$'s'} in the special relativity energy-momentum dependence as well.
We will illustrate below that an equation of this form for the wave function of a particle, even though might appear
unfamiliar to some readers, it is not uncommon in today's literature. Such an equation can be equally formulated  
for a free particle with or without a spin, a particle in an external scalar and/or in vector potential, and also in a mean field framework, 
e.g. where square roots of operators appear quite often, particularly in the formulation of GCM~\cite{Ring:2004}.

In the ground states of even-even nuclei with zero total angular momentum ${\bf P}_{coll}({\bf r})= Am{\bf V}_{coll}({\bf r})\equiv {\bf 0}$
and 
\begin{align}
\lim_{c\rightarrow\infty} K_{coll}  = \int d^3{\bf r} \frac{ mA|{\bf V}_{coll}({\bf r})|^2 }{2} =0
\end{align} 
in the case of many-fermion states in equilibrium~\cite{Bulgac:2019c}. For finite values of $c$  
the contribution to the total energy of a many-fermion system arising from the term  
$K_{coll}$ will lead to higher order corrections in $|{\bf V}_{coll}({\bf r})|^2/c^2$.  The relativistic EDF 
defined by Eq.~\eqref{eq:rDFT} in terms of various local densities,  is the Lorentzian counterpart of 
the Galilean EDF introduced more than 50 years ago by \textcite{Engel:1975}. 
$K_{coll}$ and  $K_{int} ({\bf r})$ become in the limit $c\rightarrow \infty$ the corresponding non-relativistic expressions~\cite{Bulgac:2019c,Shi:2020}
for the system and the  local intrinsic  kinetic energy densities and ${\cal E}({\bf r})$ becomes the corresponding nonrelativistic EDF.
These are  equivalents of 
\begin{align}
\varepsilon = \sqrt{ (mc^2)^2 +|{\bf p}c|^2} -mc^2, \label{eq:ek}
\end{align}
for a particle of mass $m$ and 
momentum ${\bf p}$ and of the corresponding volume elements of a fluid~\cite{LL2:1951,LL6:1959}.
The nucleon-nucleon interaction energy density $U_{pot}(\tau({\bf r}),n({\bf r}), \ldots) $  
can be parameterized as the typical non-relativistic Skyrme-like EDFs~\cite{Engel:1975}, 
and the dots are standing for any other involved number densities. 

As an illustration, for a single particle the EDF and the emerging relativistic single-particle ``Schr\"odinger'' equation for the 
single-particle wave function $\phi({\bf r})$, obtained by minimizing its total energy $\int d^3{\bf r} \, {\cal E}({\bf r})$ is
\begin{align}
&\!\!\!\! L({\bf r}) 
= \sqrt{  \bigl [ mc^2 |\phi({\bf r})|^2 \bigr ]^2 +\hbar^2 c^2 \bigl |\phi({\bf r})|^2|{\vec \nabla }\phi({\bf  r})\bigl |^2 }, \nonumber \\
&\!\!\! |{\bf p}({\bf r})c|^2 = |{\rm Re} [ -i\hbar \phi({\bf r}){\vec \nabla} \phi^*({\bf r})]c|^2  
= (\hbar^2c^2 |\phi({\bf r})|^2  |{\vec \nabla}\phi({\bf r})|^2 , \nonumber \\
&\!\!\! {\cal E}({\bf r}) = L({\bf r}) - mc^2|\phi({\bf r})|^2  , \label{eq:Sch1} \\
&\!\!\! \frac{\delta {\cal E}({\bf r})}{\delta \phi^*({\bf r})}-\varepsilon\phi({\bf r}) =
-{\vec \nabla} \cdot \left (    \frac{\hbar^2c^2 |\phi({\bf r})|^2 }{2L({\bf r})}  {\vec   \nabla} \phi({\bf r}) \right )   -\varepsilon\phi({\bf r})  \nonumber  \\
&\!\!\! +\left [ \frac{ 2(mc^2)^2|\phi({\bf r})|^2 + \hbar^2 c^2|{\vec \nabla}\phi({\bf r})|^2  }{ 2L({\bf r}) }-mc^2\right]\phi({\bf r})=0.  \label{eq:A.1}
\end{align} 
The formal structure of Eq.~\eqref{eq:A.1}, 
with a coordinate dependent effective mass, is very similar to DFT equations, where however the coordinate dependence of the effective mass
is due to the nucleon-nucleon interaction~\cite{Negele:1972}, while in this case is due to relativistic effects.  In the limit 
$c\rightarrow \infty$ the kinetic energy term in Eq.~\eqref{eq:A.1} reduces to 
\begin{align}
&-{\vec \nabla} \cdot \left (    \frac{\hbar^2c^2 |\phi({\bf r})|^2 }{2L({\bf r})}  {\vec   \nabla} \phi({\bf r}) \right )  \approx 
- \frac{\hbar^2 {\vec \nabla}^2 \phi({\bf r})}{2m}. 
 \end{align}
and $\varepsilon$, see Eq.~\eqref{eq:ek} the correct relativistic kinetic energy for free particle. In the case of a particle clearly with spin
\begin{align}
\phi({\bf r},\sigma)  =  \exp\left (  i\frac{ {\bf p}\cdot{\bf r} }{\hbar} \right )\chi(\sigma)
\end{align}
and $\psi({\bf r},\sigma)$ is also an exact solution of Eq.~\eqref{eq:A.1}
in the absence of an external potential $U({\bf r})$, where $\chi(\sigma)$ is a 2-component spinor. 
The equation Eq.~\eqref{eq:A.1} has some similarities 
with the Weyl equation for a massless 2-component spinor.  This relativistic equation however does not 
violate parity and is valid for a particle with a finite  mass and satisfies local Lorentz invariance in the time version if
$\varepsilon \rightarrow i\hbar\partial_t$. 

The relativistic EDF for a single particle in Eq.~\eqref{eq:A.1} is the exact 
equivalent of a Lorentz invariant  EDF, discarded by Dirac~\cite{Dirac:1928,Dirac:1930,Dirac:1957} for 
absence of ``symmetry between $p_0$ and $p_{1,2,3}$ required by relativity,'' which however 
was never a problem in special relativity, except when formulating a path integral for many 
fermions using Grassmann variables~\cite{Berezin:1965,Efetov:1996,Negele:1998}. This is not the case for DFT. DFT and GCM can be formulated 
in terms of Slater determinants alone~\cite{Lowdin:1955,Lowdin:1955a,Lowdin:1956,Lowdin:1956a}, 
without the explicit need of anticommuting creation and annihilation operators.  

The relativistic Schr\"odinger equation for a particle in an external potential is recovered 
with the substitution in $|({\bf r})$
\begin{align} 
mc^2\rightarrow mc^2 +U({\bf r})
\end{align}
in Eq~\eqref{eq:Sch1}.
Spin-orbit interaction can be trivially introduced as usual~\cite{Bohr:1969},  
\begin{align}
U({\bf r}) \rightarrow U({\bf r}) + ({\bf l} \cdot {\bf \sigma})\,V_{so}({\bf r}),
\end{align} 
where $ ({\bf l} \cdot {\bf \sigma})\,V_{so}({\bf r})$ is a scalar field and  Eq.~\eqref{eq:A.1} in this case becomes an equation 
for the 2-component spinor $\phi({\bf r},\sigma)$. 
This single-particle Schr\"odinger equation is non-linear, similarly to single-particle 
mean field equations, which are always nonlinear.
For a time dependent version of Eq.~\eqref{eq:A.1}  one has to perform the usual substitution $\varepsilon \phi \rightarrow i\hbar\partial_t\phi$ 
and the correct wave function 
\begin{align}
\phi({\bf r},\sigma,t)  =  \exp\left (  i\frac{ {\bf p}\cdot{\bf r} }{\hbar} -i\frac{\varepsilon t}{\hbar} \right )\chi(\sigma)
\end{align}
with the correct energy-momentum dispersion $\varepsilon = \sqrt{ (mc^2)^2 +|{\bf p}c|^2}-mc^2$ emerges.  
The lesson learned is that relativistic effects in their turn also affect the nucleon inertia in Eq.~\eqref{eq:A.1} and therefore how 
a relativistic particle responds to the presence of an external scale and/or a vector field.  
The coupling to a mean field 4-vector potential is as straightforward (minimal coupling), 
as has been implemented in the non-relativistic TDDFT~\cite{Stetcu:2015}
\begin{align} 
i\hbar \rightarrow i\hbar -A_0({\bf r}), \quad -i\hbar  \vec{\nabla}\rightarrow -i\hbar  \vec{\nabla}-\vec{A}({\bf r}).
\end{align}
The emergence of an effective mass depending nonlinearly on  the number density, 
kinetic energy number density and other densities in the single-particle
in mean field equations is a familiar occurrence 
in nuclear physics for many decades~\cite{Bender:2003,Carlsson:2008,Carlsson:2010a}.
NB The equation for a classical relativistic particle, emerging 
from the principle of least action~\cite{LL2:1951}
\begin{align}
S=-\int \!\!dt \sqrt{(mc^2)^2-|m\,\dot{\bf r}c|^2} \rightarrow \frac{d}{dt} \left ( \frac{ mc\,\dot{\bf r} }{ \sqrt{c^2-|\dot{\bf r} |^2}} \right )=0 , \label{eq:sr}
\end{align}
is also a non-linear PDE~\cite{LL2:1951}.

Since ${\bf P}_{coll}({\bf r})\equiv {\bf 0}$ the translational motion of the nucleus at equilibrium is  decoupled from the internal dynamics.  
One thus achieves a complete agreement with the physical and 
experimentally measured definition of the rest nuclear inertia
\begin{align} 
& M_Ac^2 =  \int  \!\! d^3{\bf r} \,{\cal E}_{int}({\bf r}) \label{eq:nuclear_mass}
\end{align}
The emerging mean field equations 
satisfy the generalization to Lorentzian local invariance under local gauge 
transformations, the counterpart of the Galilean local gauge invariance established  more than 50 years ago~\cite{Engel:1975}.
Relativistic extensions of other number densities, needed for odd-mass and odd-odd nuclei or for TDDFT simulations,  
can be constructed according to well known rules~\cite{Engel:1975,Dobaczewski:1995,Dobaczewski:1996}. 

One can rather easily 
convince oneself that in the case of reactions the present relativistic extension of DFT, see Eq.~\eqref{eq:rDFT}, correctly 
reproduces the relativistic energy-momentum relation~\cite{LL2:1951} for each reaction partner separately, 
as these are controlled for each reaction partner by ``its own local ${\bf P}_{coll}({\bf r})$.'' 
Thus one separates rather cleanly the local CoM motion of any small volume of matter from the corresponding intrinsic motion, as one can 
naturally achieve for CoM in the case  of a non-relativistic Galilean invariant Hamiltonian. 
The attentive reader will see here the relation between the Lagrangian and Eulerian descriptions of the hydrodynamic flow field.

Since the estimates we provide in Table~\ref{table:tab2} for 
the total kinetic energy of a nucleus are significant, a reparameterization of corresponding relativistic \`a la Skyrme EDFs, 
defined in terms of various local number densities,  is required. 
This new class of  relativistic \`a la Skyrme EDFs will not be identical to relativistic EDFs used in literature~\cite{Walecka:1974,Meng:2023}.  
In the present suggested approach, the nagging issues related to the treatment of the Dirac sea are circumvented, in 
a manner similar to other approaches in many-body theory, when only ``effective degrees of freedom'' are treated explicitly. 
After a suitable new relativistic reparametrization of the  DFT extension is achieved as outlined above the role of 
of symmetry breaking fluctuations of the mean field could be evaluated.

We have discussed above only the mean field extension of DFT.  When this DFT extension will be used for restoring various symmetries the 
EDF should be treated in a GCM framework~\cite{Ring:2004}, following Section~\ref{sec:III} and treating the expectation value 
$\langle \psi({\bf a})|\hat{\rm H}|\psi({\bf 0})\rangle$ as the corresponding EDF, expressed in mixed densities.

\section{Conclusions} \label{sec:V}

We have shown that only the  \textcite{Peierls:1957} prescription provides a better CoM energy correction
to the binding energies of nuclei in a nonrelativistic framework. It is evaluated with translationally invariant many-body wave functions, 
which are not contaminated by contributions from the excited states of a nucleus, and its average value is consistently lower by approximately 
1 MeV than the widely used CoM energy correction ${E}_{\rm CoM} =- \left \langle \psi \left | \tfrac{ \hat{\bf P}^2_{\rm CoM} }{2Am} \right |  \psi \right \rangle$. 
The \textcite{Peierls:1957}'s approach is fully in line with the other symmetry restoration methods used in literature and upon including 
relativistic effects in EDM's it will also significantly more accurately reproduce the translational nuclear inertias. Moreover, the 
suggested relativistic extension of DFT would permit the use of TDDFT in nuclear reactions with energies up to slightly 
above the pion production threshold. One would then be able to treat complex reactions at a quantum level and discuss quantum 
interference, the equilibration thermal hypothesis, 
and quantum entanglement~\cite{Bulgac:2022,Bulgac:2023,Bulgac:2024e,Bulgac:2024a,Robin:2026}. Furthermore,  one can go BMF 
and  include crucial quantum fluctuations in both static and time-dependent DFT describing non-equilibrium quantum 
dynamics~\cite{Bulgac:2024d,Kafker:2026}. Ref.~\cite{Kafker:2026} is a case in point where the translational symmetry was partially restored, 
as explained in some detail in Ref.~\cite{Bulgac:2024d}, along with the axial symmetry of the wave function describing the 
collision  $^{48}$Ca+$^{208}$Pb, and also with particle numbers of reactions fragments projections in a BMF framework.\\

{\it Acknowledgments} \textemdash  
We thank M. Bender, I. Stetcu, I. Abdurrahman,  G. A. Miller and S. Reddy for interest and discussions 
of various aspects related to this work. AB acknowledges the funding from the Department of 
Energy Office of Science, Grant No. DE-FG02-97ER41014.  This material is additionally based upon work 
supported by the Department of Energy, National Nuclear Security Administration, 
under Award Number DE-NA0004150, the Center for Excellence in Nuclear Training 
And University-based Research (CENTAUR). This research used resources of the Oak Ridge Leadership 
Computing Facility, which is a U.S. DOE Office of
Science User Facility supported under Contract No. DE-AC05-00OR22725. We are thankful as well to our last referee for inspiring us to 
clarify our stance on the need of relativistic corrections in DFT, in particular to correctly reproduce the observed nuclear translational inertia.

{\it Data availability} \textemdash The data that support the findings of
this article are not publicly available upon publication
because it is not technically feasible and/or the cost of
preparing, depositing, and hosting the data would be
prohibitive within the terms of this research project. Some amount of 
data are available from the authors upon reasonable request.

%
 % These are needed to avoid a babel error.
\providecommand{\selectlanguage}[1]{}
\renewcommand{\selectlanguage}[1]{}

\bibliography{local_fission_3R}
%{\raggedright
%\includepdf[pages={7-8]{gcm+supplement.pdf}
%}
\end{document}